\documentclass{ws-ijmpa}
\usepackage{epsfig,amsfonts,amsmath}

\def\N{{\cal N}}
\def\Z{{\mathbb Z}}
\def\C{{\mathbb C}}
\def\R{{\mathbb R}}
\renewcommand\P{{\mathbb P}}

\def\Tr{{\mbox{Tr}}}

\begin{document}

\markboth{Kristian D. Kennaway}
{Brane Tilings}

%
\catchline{}{}{}{}{}
%

\title{Brane Tilings}
\author{Kristian D. Kennaway}

\address{Department of Physics\\
University of Toronto\\
60 St. George St\\
Toronto, ON M5S 1A7, Canada}

\maketitle

\begin{history}
\received{12 June 2007}
\end{history}

\begin{abstract}
We review and extend the progress made over the past few years in understanding the structure of toric quiver gauge theories; those which are induced on the world-volume of a stack of D3-branes placed at the tip of a toric Calabi-Yau cone, at an ``orbifold point'' in K\"ahler moduli space.  These provide an infinite class of four-dimensional $\N=1$ superconformal field theories which may be studied in the context of the AdS/CFT correspondence.  It is now understood that these gauge theories are completely specified by certain two-dimensional torus graphs, called brane tilings, and the combinatorics of the dimer models on these graphs.  In particular, knowledge of the dual Sasaki-Einstein metric is not required to determine the gauge theory, only topological and symplectic properties of the toric Calabi-Yau cone.  By analyzing the symmetries of the toric quiver theories we derive the dimer models and use them to construct the moduli space of the theory both classically and semiclassically.  Using mirror symmetry the brane tilings are shown to arise in string theory on the world-volumes of the fractional D6-branes that are mirror to the stack of D3-branes at the tip of the cone.


\end{abstract}
\ccode{PACS numbers: 11.25.-w, 11.25.Uv, 11.25.Tq, 11.30.Fs}

\section{Introduction}

In recent years much of the research into the AdS/CFT correspondence has involved the detailed investigation of theories admitting fewer supersymmetries than the celebrated $AdS_5 \times S^5$/$\N=4$ Yang-Mills duality studied by Maldacena.  One approach involves deforming the $\N=4$ Yang-Mills theory by turning on relevant perturbations in the gauge theory to flow to a new CFT preserving fewer supersymmetries (typically $\N=1$).  In the gravity side this corresponds to metric and other deformations of the supergravity/string theory background (see e.g.~\cite{Lunin:2005jy}).  Another approach is to break supersymmetry by changing the topology of the string theory background, i.e.~replacing the $AdS_5 \times S^5$ geometry by a different manifold $AdS_5 \times X_5$.

In order to preserve $\N=1$ supersymmetry, $X_5$ must be a Sasaki-Einstein manifold.  One definition of a Sasaki-Einstein 5-manifold is that the metric cone over this manifold

\begin{equation}
ds ^2 = dr^2 +  r^2 dX_5^2
\label{e:cycone}
\end{equation}
is Ricci-flat, i.e.~is a Calabi-Yau 3-fold metric (the converse also holds).  In the generic $\N=1$ case the Sasaki-Einstein manifold need only admit a single $U(1)$ isometry, dual to the R-symmetry.  A seminal work in the study of the AdS/CFT correspondence for such theories is \cite{Morrison:1998cs}.  One class of Sasaki-Einstein manifolds that has proven to be very tractable are the toric Sasaki-Einstein manifolds: those for which both $X_5$ and its Calabi-Yau cone admit $U(1)^3$ isometries.

These geometries may be studied using methods of toric geometry, and there are infinitely many such toric Sasaki-Einstein manifolds, which may have complicated topology.  Thus, toric geometry provides an infinite number of $\N=1$ supersymmetric gauge theories admitting non-trivial infrared superconformal fixed points, which may be studied in the context of the AdS/CFT correspondence.

The past few years has seen the development of a new set of tools, called {\it brane tilings} and {\it dimer models}, for studying in an essentially uniform manner the structure of the four-dimensional superconformal gauge theories that are dual to the toric Sasaki-Einstein spaces.  These CFTs are quiver gauge theories admitting a $U(1)_R \times U(1)^2_F$ global symmetry, which is a subgroup of the R-symmetry group times a non-baryonic flavour symmetry group.  

Fitting into this framework are the familiar Abelian orbifolds $S^5/\Gamma$, where $\Gamma \simeq \Z_n$ or $\Z_n \times \Z_m$; the conifold and its orbifolds; and the $Y^{p,q}$ and $L^{p,q,r}$ manifolds of \cite{Gauntlett:2004yd,Cvetic:2005ft}, for all of which the Sasaki-Einstein metrics are known.  When the metric is known it allows for detailed calculations and checks of the correspondence, such as computation of correlation functions and studies of integrability.  For a generic toric manifold the Sasaki-Einstein metric is not currently known; however is now understood that for the toric theories the CFT may be {\it completely defined} without knowledge of the metric, by using the brane tiling reformulation of the gauge theory data, and the related dimer models.  In fact, as we will see, the brane tilings are realized in string theory in the context of mirror symmetry, and encode the geometry of the corresponding Calabi-Yau varieties and D-brane configurations that engineer the quiver gauge theories in string theory.

Given a particular toric Calabi-Yau 3-fold (which is necessarily a cone, as in (\ref{e:cycone})), an important problem is to determine the gauge theory that is induced on the world-volume of a stack of D3-branes at the tip of the cone\footnote{D3-branes at a non-singular point see a smooth $\C^3$ geometry locally, and this gives rise to $\N=4$ Yang-Mills in the IR, i.e.~when all curvature corrections can be neglected.} (in fact there are usually many possible theories which are equivalent in the IR, and related to one another by Seiberg duality).  These gauge theories are of quiver type and flow in the infrared to a non-trivial conformal fixed point.  As usual, this corresponds in the string theory to taking the near-horizon limit of the D3-branes, which replaces the space $\R^{3,1} \times CY_3$ by $AdS_5 \times X_5$.

The problem of determining the quiver theory for an arbitrary toric Calabi-Yau cone was initiated by \cite{Feng:2000mi}  (see \cite{He:2004rn} for a review) using the D3-brane linear sigma model of \cite{Douglas:1997de}.  The former works introduced an iterative algorithmic procedure based on embedding the singularity inside a ``larger'' (and suitably symmetric) singularity, then performing a sequence of partial resolutions of the singularity (dual to Higgsing of the symmetries of the quiver theory) until the desired endpoint is reached.  While this is a well-defined procedure for constructing the CFT in principle, it has exponential algorithmic complexity in the number of generators of the Calabi-Yau cone, and its use is limited to relatively simple singularities.

The first hint of a deeper combinatorial structure underlying the toric quiver gauge theories came with \cite{Hanany:2005ve}, which noted the existence of a set of combinatorial models called dimer models that may be used to compute the charges and field content of the D3-brane linear sigma model(s) for arbitrary toric Calabi-Yau spaces.  Making use of the dimer model structure, \cite{Hanany:2005ve} presented an improved algorithm that is more efficient than the previous computational methods for computing the D3-brane linear sigma model, but which is still somewhat limited in practise.  Shortly afterwards, the relation of this dimer model to the quiver theory was clarified and made precise in \cite{Franco:2005rj}.  Subsequent work by many authors has extended and explained the correspondence.

The structure of this review is as follows.  In section \ref{s:quiver} we define the toric quiver gauge theories that provide our class of examples, and discuss various properties of their symmetries and BPS spectrum that are relevant for the AdS/CFT correspondence.  In section \ref{s:dimer} we show how the data of the toric quiver theories may be more efficiently repackaged as an alternative structure called a brane tiling, which is a polygonal tiling of a torus.  The brane tiling is shown to encode the $U(1)$ global symmetries of the CFT via the combinatorial data of dimer configurations on the edges of the brane tiling.  We review the well-known tools from mathematical physics that solve the associated combinatorial problem, and discuss in detail how these are relevant for classifying the properties of the gauge theory.

We will show that the dimer models provide a simple construction of the moduli space of vacua of the D3-brane world-volume theory, recovering the D3-brane linear sigma models of \cite{Douglas:1997de} while avoiding the computational difficulties of that work.  We give a careful description of the quiver theory moduli space and the relation to the D3-brane linear sigma model.  In the process we find a deeper connection between the quiver theories and dimer models than had been previously understood.

It is also easy to account for the anomalous $U(1)$ symmetries of the D3-brane world-volume theory, which become massive and do not survive into the infrared to constrain the CFT.  The brane tilings then organise the classification of the scalar BPS operators of the CFT, and these operators are described by an effective theory of open and closed strings propagating on the brane tiling.  Finally one may also account for the $U(1)$ R-symmetry and its mixing with the $U(1)$ global symmetries, and implement a version of the $a$-maximization conjecture of \cite{Intriligator:2003jj} in terms of particular ``isoradial'' constraints on the embedding of the brane tiling.

Beyond their utility as a tool for studying the gauge theories, it is useful to ask how these structures arise in string theory.  In section \ref{s:geom} we discuss how the brane tilings are realised in string theory in the context of mirror symmetry.  By applying techniques from graph theory, knot theory and algebraic geometry we will show that the brane tilings encode a surprising amount of geometrical data about the mirror Calabi-Yau manifold and the geometry and topology of the mirror D6-brane configuration.  Finally section \ref{s:further} contains pointers to some other interesting topics we have not had room to address and contains some ideas for future work.

This review is largely a synthesis of the work of many authors over the past few years, although some aspects are new.  There is by now a large literature on the theory and applications of brane tilings and related systems.  I apologize to those of my colleagues to whose work I have not been able to do justice; for reasons of space I have mostly concentrated on foundational aspects in this review, and included extensive references to related work.

\section{Toric quiver gauge theories}
\label{s:quiver}

Quiver gauge theories are characterized by multiple gauge groups and matter transforming in 2-index tensor representations.   For our purposes we restrict to theories preserving $\N=1$ supersymmetry.  Quiver gauge theories are usually defined in terms of a directed graph, which specifies the representations and gauge index structure of the gauge and matter sectors.  The vertices of the graph correspond to the gauge groups; in this paper we will take the gauge groups to all be $U(N)$ (or $SU(N)$, as discussed below) for simplicity, although one may extend to the case of unequal ranks.

The directed edges of the graph (``arrows''\footnote{Hence the name ``quiver'', meaning a ``collection of arrows''}) specify the $\N=1$ chiral multiplets: an arrow between two vertices corresponds to a chiral multiplet in the bifundamental representation $(N, \bar N)$ of the $U(N) \times U(N)$ gauge groups at the two endpoints of the arrow, and an arrow from a vertex to itself corresponds to a field in the adjoint representation of that single group.   The quiver is usually considered as an {\it abstract} graph, meaning it is not considered as being embedded in any particular space, and only the connectivity of the edges is important.

The quiver graph also specifies the D-terms of the gauge theory: to each $U(1) \subset U(N)$ gauge group factor, there is an associated D-term constraint on the vacuum moduli space, whose trace is:

\begin{equation}
\sum_i Q_i^a |X_i|^2 = \zeta^a
\label{e:d-flat}
\end{equation}
where the sum is over all quiver fields $X_i$, and $Q_i^a$ is the charge of the field $X_i$ under the $a$'th gauge group $U(1) \subset U(N)$: fields in the fundamental representation of $U(N)$ have charge $+1$, those in the antifundamental $-1$, and those in the adjoint or which are uncharged under this gauge group have charge $0$ and do not participate in the D-term.  The $\zeta^a$ are Fayet-Iliopolous parameters, which are real.  The requirement of a supersymmetric vacuum is that $\sum \zeta^a = 0$.

The charge assignment agrees with the incidence matrix of the quiver, namely

\begin{equation}
 Q_i^a = \delta(a,\mbox{\it tail}(i)) - \delta(a,\mbox{\it head}(i))
\label{e:incidence}
\end{equation}
where $\mbox{\it tail}(i)$, $\mbox{\it head}(i)$ are the vertices at the two ends of the arrow $i$, and $\delta$ the Kronecker delta function.

The exact NSVZ beta function for the gauge couplings $g_a$ of the quiver gauge theory is

\begin{equation}
 \beta(g_a) = \frac{N}{1-\frac{g_a^2 N}{8 \pi^2}} \left( 3 - \frac{1}{2} \sum_{i \in a} (1 - \gamma_i)\right)
\label{ref:nsvz}
\end{equation}
where the sum is over all chiral multiplets transforming under the gauge group $a$, and $\gamma_i$ is the anomalous dimension of the field $X_i$, i.e.~the conformal dimension is

\begin{equation}
 \Delta(X_i) = 1 + \frac{1}{2} \gamma_i = \frac{3}{2} R(X_i)
\end{equation}
where $R(X_i)$ is the R-charge.  The statement of superconformal invariance is that the $\beta$-functions all vanish, which requires

\begin{equation}
\sum_{i \in a} (1 - \gamma_i) = 6 \Leftrightarrow \sum_{i \in a} (1 - R(X_i)) = 2
\label{e:scft}
\end{equation}

Not encoded in the usual presentation of the quiver graph is the superpotential, which is the remaining data needed to specify an $\N=1$ Lagrangian.  In general a superpotential must be a function of gauge-invariant operators, which correspond to closed loops on the quiver graph.  It must transform with R-charge 2 under the $U(1)_R$ symmetry, and be invariant under any global symmetries that are imposed on the theory.  Typically these are very weak constraints on the allowed form of the superpotential.

The importance of quiver gauge theories in string theory is that they arise as the world-volume theory induced on a stack of D-branes filling transverse spacetime and wrapping cycles of the internal CY space.  In the present context we consider a stack of D3-branes located at a special type of singular point of a toric Calabi-Yau cone.  The superpotential is then a definite but {\it a priori} unknown function.  

In suitably symmetric cases the superpotential is completely fixed by the requirement of invariance under the symmetries of the Calabi-Yau geometry, which act as global symmetries on the D3-branes.  Previously the determination of the superpotential by symmetry arguments was mostly limited to the gauge theories dual to $\C^3$ (i.e.~$\N=4$ SYM), the conifold \cite{Klebanov:1998hh}, and their supersymmetric orbifolds.  In a sense, the main result of the study of brane tilings is that any theory admitting a toric $U(1)^3$ global symmetry also falls into this class, and this review is a detailed explanation of how the action of this symmetry group alone is enough to completely specify the gauge theory.


What is the nature of the singularity of the toric Calabi-Yau cone on which the D3-branes are located?  It is well-known that the spectrum of BPS D-branes varies discontinuously as a function of one of the two classes of moduli of the Calabi-Yau (K\"ahler or complex structure).  In the present context, the D3-brane (which is pointlike on the Calabi-Yau) is expected to be stable at smooth points, but at singular points becomes unstable against decay into ``fractional branes'' that are localized at the singularity.  These fractional branes are only mutually BPS at special loci in the K\"ahler moduli space; for orbifold theories this is the point when orbifold symmetry is restored on the D3-brane world-volume, and the fractional branes have equal central charge.  Therefore at this ``orbifold point'' in K\"ahler moduli space they are mutually BPS and have equal mass.  In non-orbifold theories a similar picture is expected to hold: there is a locus in moduli space where the Calabi-Yau develops a conical singularity, and furthermore the fractional branes supported on this singularity become mutually BPS.  This is usually also referred to as an ``orbifold point'' even though the space may not admit an orbifold symmetry.

However, as we will see, the values of the K\"ahler moduli seen by the D3-brane world-volume theory turn out to be different.  
For D3-branes at the singularity of a Calabi-Yau cone that is at the orbifold point in moduli space, the moduli space of the world-volume quiver theory parametrizes the {\it same} Calabi-Yau at a {\it different} singular point in K\"ahler moduli space.  This is the ``conifold point'', which is where one of the fractional branes becomes massless.  This is a key difference between how closed strings and D-branes probe the geometry of spacetime.

The quiver gauge theory should be thought of as the gauge theory living on the fractional branes; this is the origin of the many gauge groups.  While it may be studied mathematically, the singular nature of the geometry tends to obscure the physics of this system of fractional branes.  Later on we will find it more convenient to invoke mirror symmetry to construct a physically equivalent D6-branes system whose geometry may be easily visualized.

Naively, one would expect the D3-brane gauge theory to consist of a product of $U(N)$ gauge groups, with an overall $U(1)$ corresponding to the center-of-mass of the D3-branes that decouples from the rest of the gauge dynamics.  Indeed, one can easily derive the quiver theory for symmetric situations such as orbifolds by imposing the orbifold projection on a parent $U(N')$ theory with $\N=4$ supersymmetry \cite{Douglas:1996sw}.

There are two subtleties about the quiver gauge theory that require careful attention.  Firstly, in the context of the AdS/CFT correspondence, the dual CFT is obtained as the IR fixed point of the D-brane world-volume theory.  The $U(1)$ gauge group factors couple to charged matter (the bifundamental chiral multiplets), and are IR free.  Therefore, at the IR fixed point the $U(1)$ gauge couplings are zero, and the gauged $U(1)$'s become {\it global} symmetries of the CFT (all matter is uncharged under the diagonal $U(1)$, and it decouples from the IR theory entirely).

The second subtle point is that some of the $U(1)$ symmetries of the classical Lagrangian are anomalous.  The anomalies are cancelled in string theory and the $U(1)$ gauge fields become massive; it is the non-anomalous $U(1)$ symmetries that survive as global symmetries in the IR.

We will refer to the D3-brane world-volume theory with $U(1)$ factors gauged (corresponding to the D3-branes on the toric CY cone) as the ``UV theory'', although it should be kept in mind that the true UV completion is the type IIB superstring theory in the presence of D3-branes.  The IR theory is obtained by going to the near horizon limit of these D3-branes, which produce an $AdS_5 \times X_5$ geometry, where $X_5$ is a Sasaki-Einstein manifold.  Type IIB string theory on this geometry is equivalent to the IR fixed point CFT, which is the $SU(N)$ quiver gauge theory; this is the statement of the AdS/CFT correspondence in this context.

$\N=1$ quiver gauge theories were first studied in string theory by Douglas and Moore \cite{Douglas:1996sw}, and later Douglas, Greene and Morrison \cite{Douglas:1997de}, who exploited the toric structure of the quiver gauge theory classical moduli space by constructing a change of variables parametrizing it as the moduli space of an Abelian gauged linear sigma model.  Many authors followed up on this work, extending the analysis to more general toric singularities and studying the properties of the resulting linear sigma models.  Of particular note is the work of \cite{Feng:2000mi} who developed algorithms for computing the quiver and superpotential for the gauge theory dual to any toric CY cone, although in practise its implementation was restricted to suitably simple theories due to computational complexity.

The quiver theories associated to D3-branes at toric singularities have additional structure that highly constrains the form of their superpotential: {\it each field appears linearly in $W$, and in precisely two terms carrying opposite sign}\footnote{This follows from the structure of the $\N=4$ Yang-Mills superpotential $W = \Tr [[X,Y],Z]$ which is fixed by supersymmetry; the  property is inherited under Abelian orbifolds, corresponding to the $\C^3/(\Z_n \times Z_m)$ singularity, and is also preserved under partial resolution/Higgsing of this theory, from which any given toric theory can be obtained for suitable $n,m$.  In this process the vevs of fields play the role of superpotential coupling constants, but it will not be important for our discussion to keep track of them.}.  In other words, all of the F-term conditions

\begin{equation}
 \frac{\partial W}{\partial X} = 0
\label{e:f-flat}
\end{equation}
 have the form {\it monomial} = {\it monomial}.  In general there may be superpotential coupling constants in addition to the opposite signs, but for now we suppress these for notational simplicity.  We will return to more general couplings in section \ref{s:geom}.

An $\N=1$ SCFT is characterized by the presence of a global $U(1)$ R-symmetry, which is dual to a $U(1)$ isometry of the Sasaki-Einstein space $X_5$.  The toric condition demands additional $U(1)$ symmetries: the isometry group is enhanced to (at least) $U(1)^3$, which acts as a combination of the R-symmetry and global flavour symmetries (there may be non-Abelian global symmetries, but we only focus on the maximal torus).  In addition there are additional $U(1)$ global ``baryonic'' symmetries in the CFT, which are gauged in the string theory on AdS.  They come from the reduction of the RR 4-form on 3-cycles of $X_5$, which produces a $U(1)$ gauge field in AdS.  We will be very explicit about these symmetries in the following section: exploiting them systematically will give rise to the dimer models, and clarify the connection to geometry and topology of the Calabi-Yau manifold.

In what follows we will mostly focus on the CFT, i.e.~the $SU(N)$ version of the quiver theory, and in particular the spectrum of scalar BPS operators; that is, the operators formed by gauge-invariant combinations of the chiral multiplets of the theory (but not the anti-chiral multiplets or the gauge multiplets).  First are the {\it mesonic} operators, which are formed by ordered sequences of operators forming a closed path on the quiver graph.  A subset of these operators are those that appear as terms in the superpotential; certain linear combinations of these operators correspond to marginal deformations of the theory that change the superpotential couplings \cite{Imamura:2007dc}.  The mesonic operators obey chiral ring relations due to the F-term constraints (\ref{e:f-flat}).

Since the CFT is based on $SU(N)$ gauge groups instead of $U(N)$, the gauge theory also admits ``di-baryonic'' operators formed by antisymmetrizing a quiver field with respect to both gauge indices:

\begin{equation}
 B_a = \epsilon_{i_1 i_2 \ldots i_N} \epsilon^{j_1 j_2 \ldots j_N} (X_a)^{i_1}_{j_1} (X_a)^{i_2}_{j_2} \ldots (X_a)^{i_N}_{j_N}
\label{e:baryon}
\end{equation}
Since we have chosen all gauge group ranks equal to $N$, these operators may be more compactly represented as determinants of the $N \times N$ matrices $X_a$:
\begin{equation}
 B_a = \det (X_a)
\label{e:det}
\end{equation}
These dibaryons are charged under the global $U(1)$ symmetries descending from the $U(1) \subset U(N)$ quiver gauge groups, with charges given by the incidence matrix of the quiver.  The FI parameter is a relevant deformation of the theory; however, the CFT spectrum is graded by baryon number(s), which play a similar role to {\it discrete} ($\Z$-valued) FI parameters \cite{Forcella:2007wk}.  We will find a very simple combinatorial way to construct the moduli space of the quiver theory and the BPS spectrum of the CFT in the next section.

There are further relations between the mesonic and baryonic operators coming from the definition (\ref{e:det}) applied to the product of fields forming a mesonic operator.  In certain cases there may also be quantum corrections to the chiral ring \cite{Beasley:2001zp}.  We will not be directly concerned with the algebraic structure of the chiral ring.  Instead, for the class of toric quiver theories, the high degree of symmetry of the vacuum allows us to use an alternative construction: we will construct the moduli space directly as a symplectic quotient, or {\it gauged linear sigma model} in physics language.  We will then see explicitly how the mesonic and baryonic operators are holomorphic functions and sections of line bundles on this space.

\section{Brane Tilings and Dimer Models}
\label{s:dimer}

In this section we begin to systematically exploit the additional structure present in the toric quiver theories to obtain some highly non-trivial simplifications.  Firstly we reformulate the quiver and superpotential as a tiling of a 2-torus by polygons (a ``brane tiling''), and then relate the set of $U(1)$ symmetries of the classical quiver theory to the combinatorics of dimer models on the brane tiling.  These dimer models naturally parametrize the moduli space of the theory as a linear sigma model.

The name ``brane tiling'' will not be justified until section \ref{s:geom} when we show that indeed all of the physics we will extract from these graphs arises from D-branes in string theory.  Specifically, we will construct a set of D-branes that engineer the quiver theories in string theory, and whose world-volume span a $T^2$ and precisely reproduce the brane tiling ``dictionary''.  For now the brane tilings will be introduced as an abstract tool which happens to provide a very compact and powerful encoding of the gauge theory data; later we will see why this is true.

If a toric quiver theory and its superpotential are known, then the corresponding brane tiling may be written down immediately, as we now discuss.  The inverse problem, namely constructing a quiver theory and its superpotential for a given toric singularity, is more subtle but can nevertheless be greatly simplified by using the connection to dimer models.  It will be studied in section \ref{s:geom}.

\subsection{Planar quivers and brane tilings}

The key idea is to regard the terms in the quiver superpotential as defining plaquettes, i.e.~forming the boundary of polygons.  When a field appears in more than one term, the plaquettes are glued together along the corresponding edge.  For the most general quiver theories the resulting object does not seem to have interesting structure, but precisely for the toric quiver theories it simplifies dramatically.  Recall that the toric quiver theories have the property that each field appears linearly in $W$, and in precisely two terms.  Thus, the plaquette tiling we form from a toric quiver theory is a polygonal tiling of a Riemann surface without boundary, called the {\it planar quiver}.

Recall that we have suppressed the superpotential couplings apart from the alternating signs $\pm 1$, translating to the rule that adjacent faces in the tiling are labelled by opposite signs.  This tiling admits an orientation (e.g.~by reversing the orientation of the boundary of alternate plaquettes), so the Riemann surface is also orientable.  What is the genus of this Riemann surface?  We can determine this from the requirement of superconformality (\ref{e:scft})

\begin{equation}
\sum_{e \in V_i} (1-R_e) = 2
\label{e:scft-v}
\end{equation}
where the sum is over edges adjacent to a vertex $V_i$, representing a gauge group of the quiver.  The requirement that the superpotential has R-charge 2 translates to
\begin{equation}
\sum_{e \in F_i} R_e = 2
\label{e:r-charge}
\end{equation}
where the sum is over edges bounding a plaquette.  Summing (\ref{e:scft-v}) over all vertices and (\ref{e:r-charge}) over all faces gives

\begin{eqnarray}
 \sum_{V_i \in V} \sum_{e \in V_i} (1-R_e) &=& 2 N_V\\
 \sum_{F_i \in F} \sum_{e \in F_i} R_e &=& 2 N_F
\end{eqnarray}
where $N_V, N_F$ are the number of vertices and faces in the planar quiver.  Since each edge connects two vertices, and separates two faces, these are equivalent to

\begin{eqnarray}
 2 \sum_{e} (1-R_e) &=& 2 N_E - 2 \sum_{e} R_e = 2 N_V\\
 2 \sum_{e} R_e &=& 2 N_F
\end{eqnarray}
where the sum is now over all edges.  Therefore $N_F - N_E + N_V = 2 - 2g = 0$; i.e.~the tiling has genus $1$ and is topologically $T^2$. 

We have succeeded in combining the (abstract) quiver graph and the superpotential to define an {\it embedding} of the quiver graph into a $T^2$, exploiting the extra symmetry that exists in the toric theories.  Furthermore, the faces may be labelled by the corresponding superpotential coupling constants, which alternate in sign between adjacent faces.

This alternating sign property is more naturally exploited by passing to the dual graph: each face is replaced by a vertex; two faces adjacent along an edge are replaced by two vertices connected by the dual edge; and the vertices are replaced by the new faces thus bounded by the dual edges.  For a graph on $T^2$ this dual graph may always be uniquely constructed and also lives on a $T^2$.

The alternating sign property satisfied by the vertices is the definition of a {\it bipartite} graph: each vertex is only connected to vertices of the opposite sign, or ``colour''; from now on we will mostly refer to the positve and negative vertices as black and white.  This bipartite property enjoyed by the toric quiver theories is the key fact that we systematically exploit for the rest of this section.

The dictionary between the quiver gauge theory, planar quiver and brane tiling is summarized in figure \ref{t:dictionary}.

\begin{figure}[tb]
\begin{center}
\begin{tabular}{l|l|l}
Gauge theory & Planar quiver & Brane tiling\\\hline
$SU(N)$/$U(N)$ gauge group & Vertex & Polygonal face\\
Chiral multiplet & Edge & Edge\\
Superpotential term & Polygonal face & Vertex
\end{tabular}
\end{center}
\caption{The quiver gauge theory/planar quiver/brane tiling dictionary.\label{t:dictionary}}
\end{figure}

\subsection{Example: Brane tiling for del Pezzo 1}
\label{s:dp1-tiling}

\begin{figure}[tbh]
\begin{center}
 \epsfig{file=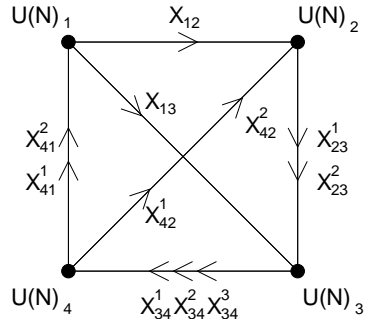}
\end{center}
 \caption{The quiver for D3-branes at a conical del Pezzo 1 singularity.\label{f:dp1-quiver}}
\end{figure}

\begin{figure}[tbh]
\begin{center}
 \epsfig{file=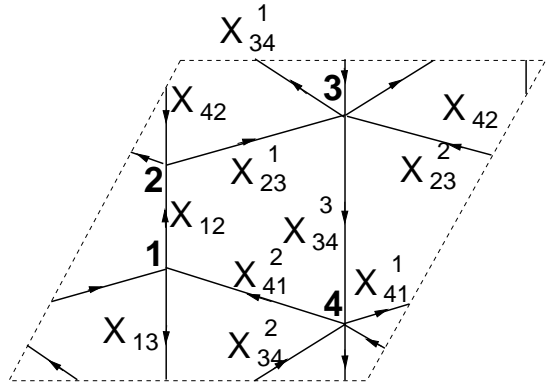,height=2in}
\end{center}
 \caption{The planar quiver for the del Pezzo 1 quiver.\label{f:dp1-planar}}
\end{figure}

\begin{figure}[tbh]
\begin{center}
 \epsfig{file=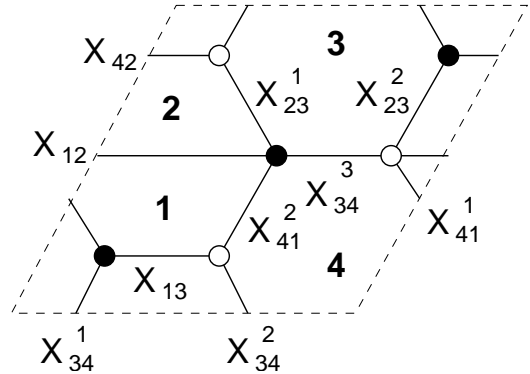}
\end{center}
 \caption{The brane tiling for the del Pezzo 1 quiver.\label{f:dp1-tiling}}
\end{figure}

The quiver for D3-branes at a conical del Pezzo 1 singularity is shown in figure \ref{f:dp1-quiver} \cite{Feng:2000mi}.  The superpotential is

\begin{eqnarray}
 W &=& \Tr X_{13} X_{34}^2 X_{41}^2 - \Tr X_{13} X_{34}^1 X_{41}^1 + \Tr X_{12} X_{23}^2 X_{34}^3 X_{41}^1 \\ \nonumber
&&- \Tr X_{12} X_{23}^1 X_{34}^3 X_{41}^2 + \Tr X_{23}^1 X_{34}^1 X_{42} - \Tr X_{23}^2 X_{34}^2 X_{42}
\end{eqnarray}
The quiver and superpotential may be combined into the planar quiver shown in figure \ref{f:dp1-planar}.  The brane tiling is shown in figure \ref{f:dp1-tiling}.

\subsection{Quiver gauge theories and dimer models}
\label{s:quiver-dimer}

We have seen how for a toric quiver theory, the data of the matter content (quiver) and interactions (superpotential) may be naturally combined into a single object, the planar quiver, or equivalently its dual graph, called a {\it brane tiling}.  We will see that the global symmetries of a toric quiver theory are enough to completely fix its classical moduli space.  In particular we focus on the $U(1)$ global symmetries (the maximal torus if the global symmetry group contains non-Abelian factors).  The problem of enumerating the $U(1)$ symmetries translates into a combinatorial problem on the brane tiling, and we are led to study the combinatorics of {\it dimer models}.  The dimer models produce a linear sigma model description of the gauge theory classical moduli space, precisely reproducing the construction of \cite{Douglas:1997de}.

However, the linear sigma model for the classical moduli space is not the end of the story, because some of the $U(1)$ classical symmetries are broken by anomalies in the quantum theory.  Remarkably, there is also an easy way to read off the non-anomalous symmetries from the dimer model, and we may also use the dimer model to construct a linear sigma model whose gauge group is the group of anomaly-free symmetries.  We first discuss the classical theory and return to the question of anomalies in the next subsection.

\subsubsection{The classical $U(1)$ symmetries of the toric quiver theories}

We start by looking for the set of $U(1)$ actions (i.e.~phase rotations) on the quiver fields $X_{ij}^k$ under which the superpotential transforms homogeneously.  Note that the kinetic and D-terms are automatically invariant under this operation since they pair each field $X_{ij}^k$ with its complex conjugate.  Once we know the homogeneous transformations of the superpotential, we can form $U(1)$ symmetries of the Lagrangian by taking the quotient of two such $U(1)$ actions with appropriate weights so that the superpotential is invariant.  This problem has a beautiful combinatorial description, which we now describe.

Given a bipartite graph, a {\it dimer} is a marked edge, which by definition connects a black and white vertex.  A {\it dimer configuration} or {\it perfect matching} is a collection of dimers chosen so that every vertex of the graph is covered by exactly one dimer.  The bipartite graph together with its set of dimer configurations will be called a {\it dimer model}.  The dimer models have a long history in condensed matter, chemistry and mathematical physics.

Under the dictionary we have presented, the vertices of the graph are the superpotential terms of the quiver Lagrangian.  A dimer configuration is therefore a choice of exactly one field in every term of the superpotential, which furthermore appears linearly.  Therefore, acting by common phase rotations on this set of fields acts homogeneously on $W$.  We have translated the problem of finding the set of homogeneous transformations of $W$ into the problem of enumerating the perfect matchings of the bipartite graph.  This problem was solved by Kasteleyn \cite{Kasteleyn} in 1967.  We will use a modification of the related form of the Kasteleyn construction given in \cite{Kenyon:2003uj}.

Given the bipartite graph $\Gamma$, we form the {\it weighted adjacency matrix} $A$, whose rows and columns index the black and white vertices, and whose entries are

\begin{equation}
 A_{ij} = \left\{ \begin{array}{ll}
                  	\sum_k a_{ij}^k& \mbox{for each edge $k$ connecting black vertex $i$ to white vertex $j$}\\
			0 & \mbox{otherwise}
                  \end{array}\right.
\end{equation}
where the $a_{ij}^k \in \R^*$ are (for now) formal variables labelling the edges, called {\it edge weights}.  This matrix $A$ specifies the connectivity of $\Gamma$, which completely determines it as an abstract graph (i.e.~forgetting about its embedding into the torus).  This is already enough to determine its set of matchings, but we will make use of a refinement that keeps track of the embedding $\Gamma \subset T^2$.

To define the {\it Kasteleyn matrix}, choose a representative of the two primitive winding cycles of the torus, called $\gamma_w$ and $\gamma_z$.  We can choose these paths to be the boundary of the fundamental domain of the $T^2$, but any independent choice will suffice.  Different choices for the homology classes of $\gamma_w$ and $\gamma_z$ as well as their explicit representatives will turn out to induce certain linear transformations that are physically irrelevant.  Then the Kasteleyn matrix $K(w,z)$ is defined by
\begin{equation}
 K_{ij} = \sum_k a_{ij}^k z^{<a_{ij}^k, \gamma_z>} w^{<a_{ij}^k, \gamma_w>}
\end{equation}
where $<a_{ij}^k, \gamma>$ is the (signed) intersection number of the edge represented by $a_{ij}^k$ (with the natural bipartite orientation) and the oriented contour $\gamma$.  It takes values $\pm 1, 0$, if the edge crosses $\gamma$ with positive or negative orientation, or does not cross $\gamma$.

The Kasteleyn matrix is the basic object we will need to recover the D3-brane linear sigma models, which describe the classical and semiclassical (i.e.~anomaly-free) moduli space of the quiver gauge theory.  The utility of the Kasteleyn matrix is that its determinant enumerates the perfect matchings of the bipartite graph\footnote{With an addition rule for assigning $\pm$ signs to the entries of $K$ \cite{Kasteleyn,Kenyon:2003uj} it is possible to arrange for all of the coefficients of $z^a w^b$ to have the same sign, so that setting all $a_{ij}^k = 1$ counts the matchings without cancellations.  We will return to this in section \ref{s:iso}.}.  It is easy to see that the cofactor expansion of the determinant precisely reproduces the definition of the dimer configurations given above: the rows and columns of $K$ define the vertices of the graph (superpotential terms), and the cofactor expansion selects precisely one edge from each row and column of the matrix, in all possible ways.  The reason for including the factors of $z, w$ in the Kasteleyn matrix is that the determinant expansion has an interesting bi-grading by the exponents $z^a w^b$, which provides a connection to toric geometry, as we will see.  

\subsubsection{Example: Dimer model for del Pezzo 1}
\label{s:dp1}

The Kasteleyn matrix for the del Pezzo 1 brane tiling of figure \ref{f:dp1-tiling} is (choosing an arbitrary ordering of the vertices)

\begin{equation}
 K(w,z) = \left(\begin{array}{cccc}
                a_{13} & a_{41}^1 z & a_{34}^1 \frac{1}{w}\\
		a_{41}^2 & a_{34}^3 + a_{12} z& a_{23}^1 \\
		a_{34}^2 w & a_{23}^2 & a_{42} \frac{1}{z}
                \end{array}\right)
\label{e:dp1-det}
\end{equation}
with determinant

\begin{eqnarray}
 det K(w,z) &=& (-a_{13} a_{23}^1 a_{23}^2 - a_{34}^1 a_{34}^2 a_{34}^3 + a_{12} a_{13} a_{42} - a_{41}^1 a_{41}^2  a_{42}) \\\nonumber
&& + a_{23}^2 a_{34}^1 a_{41}^2 \frac{1}{w} + a_{13} a_{34}^3 a_{42} \frac{1}{z} - a_{12} a_{34}^1 a_{34}^2 z + a_{23}^1 a_{34}^2 a_{41}^1 w z
\end{eqnarray}
The summands correspond to the 8 perfect matchings and are shown in figure \ref{f:dp1-matchings}.

For many other examples of dimer models corresponding to toric singularities, see \cite{Hanany:2005ve,Franco:2005rj}.

\subsection{Review of toric geometry}
\label{s:toric}

The standard mathematical reference on toric geometry is \cite{Fulton}.  We quickly review some key concepts and notation.

A {\it toric variety} is defined by an integer lattice $N \simeq \Z^n$ ($n=3$ for our purposes), and a {\it fan} of strongly convex rational polyhedral cones generated by elements of $N$; these are cones with apex at the origin, generated by elements of $N$ as a vector space over $\R$.  It is most convenient for us to define the toric variety as a symplectic quotient, or gauged linear sigma model.  To each lattice generator $N_i \in N$, we associate a coordinate $z_i \in \C$.  The $d-n$ linear relations between the $d$ generators $N_i$ of $N$ may be parametrized by integer ``charges'' $Q$,

\begin{equation}
 \sum_{i=1}^{d} Q_i^{(a)} N_i = 0
\end{equation}
which determine a $(\C^*)^{d-n}$ action on the $z_i$, via

\begin{equation}
 (z_1, z_2, \ldots z_{d}) \sim (\lambda_a^{Q_1^{(a)}} z_1, \lambda_a^{Q_2^{(a)}} z_2, \ldots \lambda_a^{Q_d^{(a)}} z_d)
\label{e:toric}
\end{equation}
where $\lambda_a \in \C^*$, $a=1, \ldots, d-n$.  The toric variety is obtained by performing a symplectic quotient $\C^{n+d} // U(1)^d$, which is defined by first imposing the equations

\begin{equation}
 \sum_{i=1}^{d} Q_i^{(a)} |z_i|^2 = \mu_i
\label{e:level}
\end{equation}
and then dividing by the $U(1)^a$ symmetry corresponding to the angular part of (\ref{e:toric}).  The real parameters $\mu_i$ are called the levels of the $i$'th moment map.  Note the similarity to the D-term constraints (\ref{e:d-flat}) of a gauged $U(1)^a$ symmetry with FI parameters $\mu_i$; this is the connection between toric geometry and the linear sigma models.

When all of the lattice vectors $N$ are coplanar, then the resulting variety is Calabi-Yau.  Equivalently, each of the charge vectors $Q^{(a)}$ satisfies

\begin{equation}
 \sum_{i=1}^{d} Q_i^{(a)} = 0
\end{equation}
By performing an $SL(3,\Z)$ transformation on the lattice $N \simeq \Z^3$, they may be brought to the form $(1,a,b)$ where $(a,b)$ span a convex integer polygon $\Delta \subset \Z^2$, called the {\it toric diagram}.  We restrict to the Calabi-Yau 3-fold case from now on.

Given a choice of levels $\mu_i$, equations (\ref{e:level}) present the toric Calabi-Yau as a $U(1)^3 \simeq T^3$ bundle over a real linear subspace (parameterized by the $|z_i|^2$ variables) formed by the intersection of the hyperplanes (\ref{e:level}) in $(\R^+)^{d}$.  For generic choices of $\mu_i$ the resulting space is smooth.  When all $\mu_i = 0$ the origin $(z_1, \ldots, z_d) = 0$ is a solution of (\ref{e:level}) and the Calabi-Yau space is a complex cone over a toric surface, with a Gorenstein canonical singularity at the origin.  Intermediate choices of $\mu_i$ produce partial resolutions of this singularity.

\begin{figure}[t]
 \begin{center}
 \epsfig{file=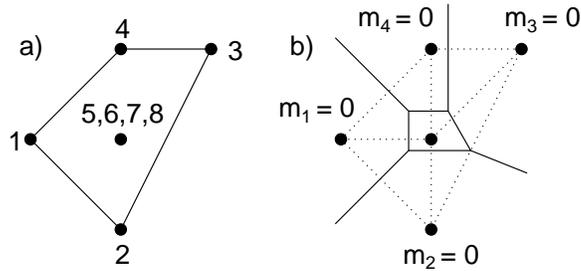,width=3in}
\end{center}
\caption{a) The Newton polygon $\Delta$ for the toric del Pezzo 1 surface, with the labels corresponding to perfect matchings.  b) A complete triangulation of the toric diagram, corresponding to the resolution of the singular del Pezzo cone, and the dual $(p,q)$ web.\label{f:toric}}
\end{figure}

This real subspace may be easily visualized by projecting onto a transverse 2-dimensional plane.  The 1-dimensional boundaries project to lines in $\R^2$, which are often called a ``$(p,q)$ web'' in the physics literature\footnote{In a dual picture the edges of the web correspond to 5-branes carrying $(p,q)$ D5- and NS5- charge \cite{Aharony:1997bh}}.  This graph is dual (as a planar graph) to a triangulation of the toric diagram; see figure \ref{f:toric}.  Each of the non-compact boundary components of (\ref{e:level}) is characterized by the vanishing of one of more of the coordinates $z_i$; in particular the top-dimensional boundary components are defined by $z_i = 0$.  These define a codimension 1 subspace of the complex 3-fold, and give non-compact 4-dimensional submanifolds called toric divisors.  By the planar duality, the toric divisors are 1-1 with the lattice generators on the boundary of the toric diagram; this fact will be very important later on.

The toric divisors play an important role in the AdS/CFT correspondence: as we have mentioned, the singular Calabi-Yau manifold is also equivalent to a {\it real} cone over a real 5-manifold; this is one definition of the 5-manifold to be Sasaki-Einstein.   When the Calabi-Yau is toric, this Sasaki-Einstein manifold is itself a $U(1)$ bundle over a toric surface, and admits a $U(1)^3$ isometry.  In this case, the non-compact 4-dimensional toric divisors restrict to {\it compact} 3-cycles in $X_5$.  These 3-cycles correspond to toric divisors of the toric surface (which are holomorphic curves), together with the $U(1)$ fibre over them.  These 3-cycles will be crucial for understanding the non-anomalous $U(1)$ symmetries of the quiver theories in section \ref{s:anom}.

\subsection{Dimer models and toric geometry}
\label{s:dimers-toric}

\begin{figure}[p]
\begin{center}
 \epsfig{file=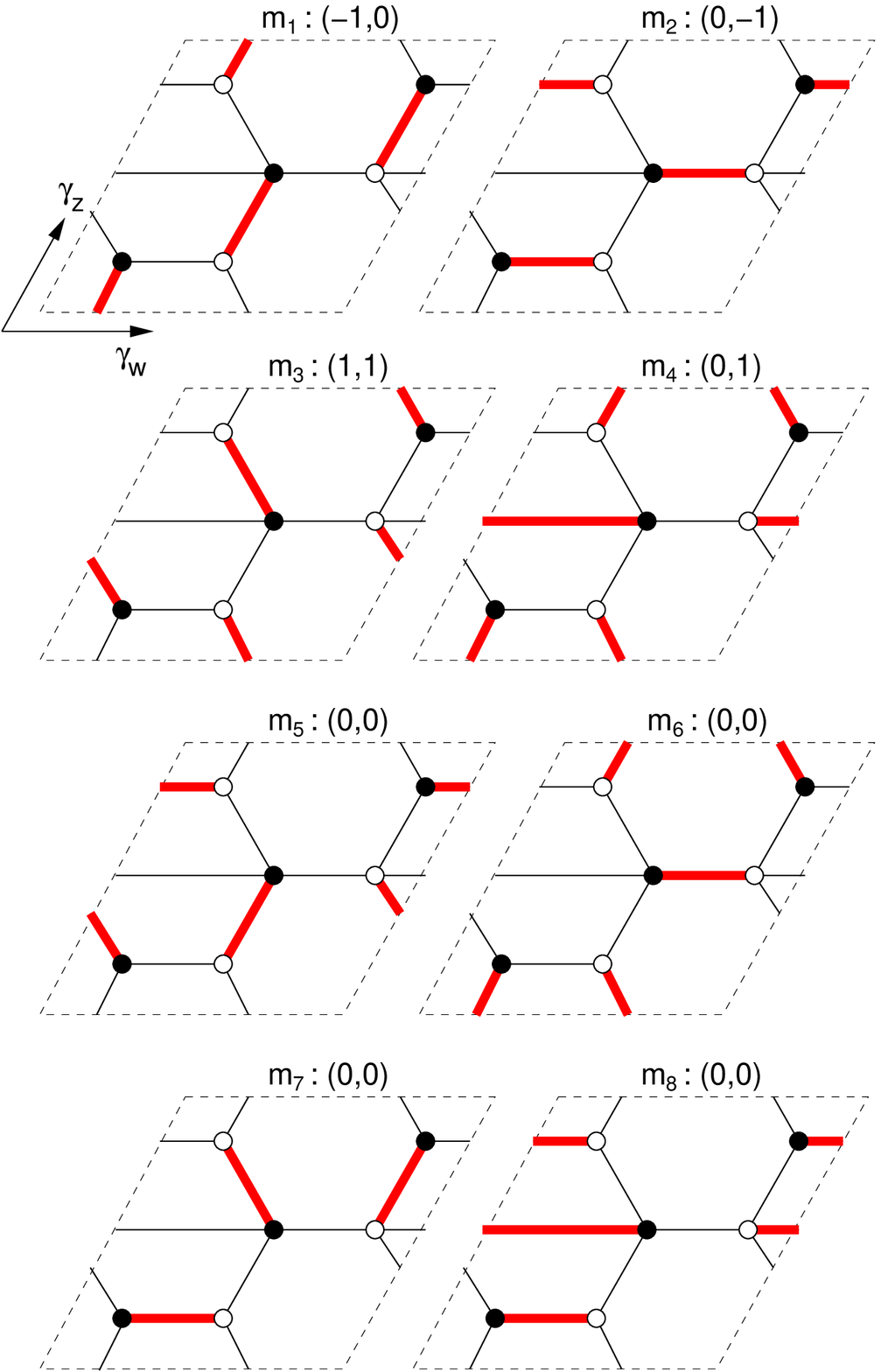,height=7in}
\end{center}
 \caption{The 8 perfect matchings of the del Pezzo 1 quiver, with weights $(h_w, h_z)$ prescribed by the intersection of the matching with the paths $\gamma_w, \gamma_z$, which we choose to be the $(1,0)$ and $(0,1)$ cycles at the boundary of the fundamental domain.\label{f:dp1-matchings}}
\end{figure}

The determinant of the Kasteleyn matrix may be expanded as

\begin{equation}
 det K(w,z) = \sum_{(a,b) \in \Delta} f_{ab}(a_{ij}^k) w^a z^b
\end{equation}
where $\Delta$ enumerates the set of exponents, and $f_{ab}(a_{ij}^k)$ is a polynomial of the edge weights $a_{ij}^k$ that is {\it linear} when written in terms of the matchings, i.e.~we may define the matching variables $m_i$ so that

\begin{equation}
 f_{ab}(a_{ij}^k) = \sum_{l \subset M_{ab}} m_l
\end{equation}
where the sum runs over a subset of the matchings $M_{ab} \subset M$.  The assignment of matchings to terms in the expansion of $\det K$ is determined by their intersection with the paths $\gamma_w, \gamma_z$, as above.  See Figure \ref{f:toric} for the Newton Polygon of $\det K(w,z)$ for the del Pezzo 1 quiver studied in section \ref{s:dp1}.  The integers $i$ label the matchings $m_i$ at each of the lattice points $(a,b) \in \Z^2$, i.e.~coefficients of $w^a z^b$ in $\det K$ (compare figure \ref{f:dp1-matchings}).

Given two matchings $m_1, m_2$ of the graph $\Gamma$, their difference $m_1 - m_2$ gives a collection of oriented, closed curves on $\Gamma$.  The winding number $(p,q)$ is related to the lattice weights $w^a z^b$ via the intersection form on $T^2$,

\begin{equation}
 (p_1, q_1) \cap (p_2, q_2) = \det\left(\begin{array}{cc}
                                        p_1 & p_2 \\
					q_1 & q_2
                                      \end{array}\right)
\end{equation}
For any two matchings with lattice points $w^a z^b, w^c z^d$, the curve produced by their difference $m_1 - m_2$ will have homology class on $T^2$ given by

\begin{equation}
-(b-d) [\gamma_w] + (a-c) [\gamma_z] 
\end{equation}
i.e.~winding number $(d-b,a-c)$ if the contours $\gamma_w, \gamma_z$ are taken to be the boundary of the fundamental domain.  

Given $det K(w,z)$, form the Newton Polygon $\Delta$, by taking the convex hull of the exponents $(a,b) \in \Delta$, thought of as lattice points in $\Z^2$.  The observation of \cite{Hanany:2005ve} is that this lattice polygon is closely related to the toric geometry of the Calabi-Yau space on which the D3-branes were originally defined\footnote{Later, we will take this as part of the definition of the class of bipartite graphs that are relevant for describing quiver gauge theories.}.  Indeed, if we extend the lattice vectors

\begin{equation}
(a,b) \mapsto (1,a,b)
\end{equation}
then these are precisely the generators of the toric fan of the (resolved) Calabi-Yau cone on which we placed the stack of D3-branes, and $\Delta$ is the toric diagram of the Calabi-Yau.  Different choices of the paths $\gamma_w, \gamma_z$ used to define $K(w,z)$ act by $SL(3,\Z)$ transformations on the set of lattice vectors, which act on $\Delta$ by $SL(2,\Z)$ and an integer shift, and by a linear change of basis of the charge vectors $Q$.  Clearly these basis changes do not change the underlying Calabi-Yau geometry.

The reason for the correspondence between dimer models and toric Calabi-Yau geometry will become clear in the following section.  The basic idea is that the D3-branes are probe branes on the Calabi-Yau geometry, so they are free to explore the cone geometry and their moduli space will reflect the freedom for the D3-branes to explore the neighbourhood of the singularity.  The dimer model will give a natural parametrization of this moduli space that exploits its toric structure.

One important characteristic of the dimer models is that the lattice vectors $(1,a,b)$ often have multiplicity greater than one; this is the statement that the coefficient of $z^a w^b$ in $\det K$ may have several summands.  Thus, they are not a ``minimal'' presentation of the Calabi-Yau geometry, but this turns out to have an interesting physical implication.

\subsection{Preferred bases of matchings}
\label{s:basis}

Given the set of matchings associated to a dimer model, any pairwise differences of them generate a $U(1)$ symmetry of the theory.  The differences of matchings also correspond to closed, oriented paths on the edges of the bipartite graph.  
We will classify these symmetries according to the homology of these paths, and identify certain distinguished bases for the symmetries.

We may classify the various $U(1)$ symmetries more precisely by recasting them as differential forms on the graph \cite{Imamura:2006ie}.  A 1-form $\epsilon$ is a function on the edges of the graph
\begin{equation}
 \epsilon = \sum_{i=1}^{N_E} \epsilon_i \delta(i)
\end{equation}
where $\delta(i)$ is the delta-function supported along the $i'th$ edge with its canonical (black $\rightarrow$ white) orientation.  The 1-form $\epsilon$ is antisymmetric under change of orientation

\begin{equation}
 \epsilon(-e_i) = - \epsilon(e_i)
\end{equation}
It is defined up to a gauge transformation

\begin{equation}
\epsilon \sim \epsilon + df
\label{e:gauge} 
\end{equation}
where the differential is defined by
\begin{equation}
 df(e) = f(F_L) - f(F_R)
\label{e:d0}
\end{equation}
i.e.~each edge $E$ on the graph receives a contribution from the difference of an (integer valued) function evaluated on the faces to the left and right of the edge, with respect to the bipartite orientation of the graph (i.e.~looking from the black to the white vertex).

An assignment of $U(1)$ charges to the quiver fields corresponds to a 1-form on the edges of the graph.  We require that the superpotential be uncharged with respect to all $U(1)$ symmetries\footnote{The $U(1)_R$, under which $W$ has charge 2, will be treated separately in section \ref{s:iso}.}, which translates to the closure condition.

\begin{equation}
 d\epsilon = 0
\end{equation}
Therefore the $U(1)$ symmetries are classified by the elements of $H^1(\Gamma; \Z)$.

\subsubsection{Baryonic and mesonic symmetries}
\label{s:symmetries}

A mesonic operator $M$ corresponds to a closed path $\gamma_M$ on $T^2$ that crosses a sequence of edges.  Therefore the charge of this operator with respect to a 1-form is given by the contour integral,

\begin{equation}
 Q_\epsilon(M) = \oint_{\gamma_M} \epsilon
\label{e:contour}
\end{equation}
which receives contributions from the edges crossed by $\gamma_M$, which is where $\epsilon$ is supported.

By definition, the baryonic $U(1)$ symmetries are those under which all mesons are uncharged.  Therefore (\ref{e:contour}) vanishes on all contours, and the 1-form $\epsilon$ is exact.  A basis of $H^0(\Gamma, Z) \simeq \Z^{F-1}$ is given by the delta functions $\delta({F_i})$ supported on the faces of the tiling.
With respect to the $i$'th face, $\delta(F_i)$ produces a charge assignment $d \delta(F_i)$ on the edges (via the rule (\ref{e:d0})) that is alternately $\pm 1$ around the boundary of the face (since the orientation of the edge, and therefore the sign of the charge, alternates), with the exception that the charge is $0$ if the edge is adjacent to another copy of the same face by periodicity of the torus.

It is easy to see that this is precisely the charge assignment of the incidence matrix of the quiver (\ref{e:incidence}), and agrees with the ``baryonic'' $U(1)$ charges of all the fields in the CFT. 
%
%
By definition, since a mesonic operator is a closed path on the quiver, it must enter a gauge group as many times as it leaves.  Therefore all mesonic operators are uncharged under these $U(1)$'s; on the other hand, the baryons formed from fields incoming or outgoing to the gauge group are charged.  This justifies the name ``baryonic'' for these $U(1)$ symmetries in the CFT limit.  For later use we will also refer to this parametrization of these symmetries as ``face''  symmetries, since they come from contours surrounding a face of the tiling.

What about the mesonic symmetries?  There are precisely two non-trivial generators of 1-forms on the graph, corresponding to paths winding the two nontrivial cycles of the torus.  These are the mesonic $U(1)$ flavour symmetries, under which the mesons are charged (i.e.~some of the paths corresponding to mesons will intersect with these contours).  Of course, they may mix with the baryonic symmetries, corresponding to a gauge transformation (\ref{e:gauge}).  It is known that the mesonic ``flavour'' symmetries correspond geometrically to $U(1)^2$ isometries of the toric surface at the base of the toric CY cone, and this agrees nicely with their parametrization as winding cycles of the $T^2$.  Indeed, we will see via mirror symmetry in section \ref{s:geom} that the $T^2$ of the brane tiling is identified with this torus of isometries.

\subsubsection{Representation as difference of matchings}
\label{s:diff-match}

Starting from the Kasteleyn matrix, form the matching matrix

\begin{equation}
 M = (M)_{i \alpha} = \left\{
\begin{array}{ll}
 1 & \mbox{edge $i$ is contained in matching $\alpha$}\\
 0 & \mbox{otherwise}
\end{array}\right.
\end{equation}
In practise this matrix can be read off from the coefficients of the terms in $\det K$.  As we discussed, the columns of $M$ give a parametrization of the set of $U(1)$ transformations under which the superpotential scales homogeneously, and differences of columns generate the $U(1)$ symmetries of $W$.

By the discussion in the previous section, it is clear that the charge assignments formed by differences of columns of $M$ agree with the parametrization in terms of 1-forms.  In particular, there is a preferred basis of matchings that generate the contours surrounding the faces, and these are the classical baryonic symmetries.  By the  argument of section \ref{s:dimers-toric} that maps matchings to lattice vectors, these symmetries must correspond to sums of differences of matchings whose lattice vectors $(1,a,b)$ sum to zero, and the corresponding contour encircles a face and has winding $(0,0)$ on $T^2$.

Finally, it is usually the case that the parametrization of homogeneous transformations of $W$ by matchings contains some redundancies: the perfect matchings are not all independent, but satisfy identities.  In other words, the perfect matchings often form an over-complete basis for the space of $U(1)$ symmetries; we will need to divide out by the resulting equivalences (which act trivially on the quiver fields).  Technically, this can be implemented by gauging the corresponding ``trivial'' $U(1)$, but it should be remembered that this does not correspond to an actual gauge symmetry of the original quiver theory.  On the other hand, since all quiver fields are uncharged under such $U(1)$'s, we could choose to add them to the quiver theory and they would be completely decoupled from the original theory.

\subsubsection{Zig-zag symmetries}

There is an alternative set of 1-forms that will turn out to be very useful for accounting for the non-anomalous $U(1)$ symmetries of the quiver theory.  These are generated by the differences of matchings that are boundary points of $\Delta$; we may choose the anti-clockwise ordering of points on the boundary and take successive pairwise differences $m_{i} - m_{i+1}$.  Clearly these are all paths with non-trivial winding number $(p,q)$, but they are subject to various linear relations among the vectors $(p,q)$.  They are therefore equivalent to certain linear combinations of the baryonic and mesonic symmetries discussed above.

These $(p,q)$ winding paths are called ``zig-zag paths'', for reasons that will become clear in section \ref{s:graph}.  These zig-zag paths will be of central importance for understanding the geometry of the dimer models via mirror symmetry.  For now, we may note that since they are defined by differences of the external matchings, they are closely related to the toric geometry of the singular Calabi-Yau cone (they define differences of toric divisors).

\begin{figure}[t]
 \begin{center}
 \epsfig{file=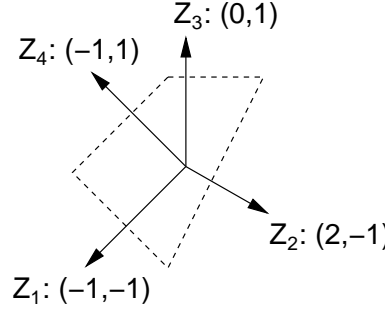}
\end{center}
\caption{The outward-pointing normal vectors to the Newton polygon correspond to the winding numbers of the zig-zag paths.\label{f:dp1-normal}}
\end{figure}

Note that the winding number (homology class) of the zig-zag paths always correspond to the outward-pointing normal vectors of the Newton polygon, see Figure \ref{f:dp1-normal}.  This follows from the relation between lattice vectors and winding numbers in section \ref{s:dimers-toric}.

\subsubsection{Example: Charges for del Pezzo 1}
\label{s:dp1-charges}

From (\ref{e:dp1-det}), we read off the matching matrix

\begin{equation}
 M = \left(\begin{array}{c|cccc|cccc}
		         &m_1&m_2&m_3&m_4&m_5&m_6&m_7&m_8\\\hline
		X_{12}   & 0 & 0 & 0 & 1 & 0 & 0 & 0 & 1 \\
		X_{23}^1 & 0 & 0 & 1 & 0 & 0 & 0 & 1 & 0 \\
		X_{23}^2 & 1 & 0 & 0 & 0 & 0 & 0 & 1 & 0 \\
		X_{34}^1 & 1 & 0 & 0 & 1 & 0 & 1 & 0 & 0 \\
		X_{34}^2 & 0 & 0 & 1 & 1 & 0 & 1 & 0 & 0 \\
		X_{34}^3 & 0 & 1 & 0 & 0 & 0 & 1 & 0 & 0 \\
		X_{42}   & 0 & 1 & 0 & 0 & 1 & 0 & 0 & 1 \\
		X_{13}   & 0 & 1 & 0 & 0 & 0 & 0 & 1 & 1 \\
		X_{41}^1 & 0 & 0 & 1 & 0 & 1 & 0 & 0 & 0 \\
		X_{41}^2 & 1 & 0 & 0 & 0 & 1 & 0 & 0 & 0 \\
           \end{array}\right)
\end{equation}
where we have split the matchings into external points on the toric diagram ($m_1, \ldots, m_4$) and internal points ($m_5, \ldots, m_8$); compare Figure \ref{f:toric}.

The face symmetries are given by the following linear combination of matchings
\begin{equation}
 Q_D=\left(\begin{array}{c|cccccccc|c}
		F_1 & 0 & 0 & 0 & 0 & -1 & 0 & 0 & 1& \zeta^1\\
		F_2 & 0 & 0 & 0 & 0 & 0 & 0 & 1& -1 & \zeta^2\\
		F_3 & 0 & 0 & 0 & 0 & 0 & 1& -1 & 0 & \zeta^3\\
		F_4 & 0 & 0 & 0 & 0 & 1& -1 & 0 & 0 & \zeta^4\\
           \end{array}\right)
\end{equation}
where we have also indicated the FI parameters $\zeta^a$.  In this case the face symmetries only depend on the internal matchings, but this is not true in general.  Note that only 3 of the face symmetries are independent, and we may eliminate $F_4$ as redundant via $\sum \zeta^a = 0$.

The zig-zag symmetries are defined to be

\begin{equation}
 Q_Z=\left(\begin{array}{c|cccccccc}
		Z_1 & 1 & -1 & 0 & 0 & 0 & 0 & 0 & 0\\
		Z_2 & 0 & 1 & -1 & 0 & 0 & 0 & 0 & 0\\
		Z_3 & 0 & 0 & 1 & -1 & 0 & 0 & 0 & 0\\
		Z_4 & -1 & 0 & 0 & 1 & 0 & 0 & 0 & 0\\
           \end{array}\right)
\label{e:dp1-qz}
\end{equation}
Again one of them (say $Z_4$) is linearly dependent on the others.  We will use this basis of symmetries in section \ref{s:anom} to construct the anomaly-free baryonic symmetries.

Finally, the redundancies are given by

\begin{equation}
 Q_F = \mbox{ker} M = \left( \begin{array}{cccccccc|c}
		0 & 1 & 0 & 1 & 0 & -1 & 0 & -1 &0\\
		1 & 1 & 1 & 0 & -1 & -1 & -1 & 0 &0\\
                        \end{array}\right)
\end{equation}
Here the right-most column records the fact that there are no FI parameters corresponding to these symmetries, since they do not correspond to gauge groups of the quiver Lagrangian.

\subsection{The classical D3-brane linear sigma model in detail}
\label{s:classical}

We consider first the UV theory with gauge groups $U(N)$ and corresponding D-terms (\ref{e:d-flat}), in the classical limit.  The space of solutions to the F-term equations (\ref{e:f-flat}) spans an affine toric cone; similarly the D-term equations (\ref{e:d-flat}) also span a non-compact toric variety (it may be non-singular if the FI parameters are non-zero).  We may parametrize the simultaneous solutions to the F- and D-term constraints (the intersection of these two toric spaces, which is again toric) as a symplectic quotient, or gauged linear sigma model.  Constructing this ``D3-brane linear sigma model'' is the goal of this section.

This construction was first presented in \cite{Douglas:1997de} and subsequently studied and extended by many authors.  However these older treatments, while mathematically well-defined, did not exploit the full symmetry of the problem and a key step in the construction (the computation of a certain dual cone) has exponential computational complexity, which limited the application of the techniques in practise.  We will instead give an alternative construction that takes into full account the combinatorial structures we have exposed.  The moral is that the dimer configurations are the natural variables that were missing from the construction of \cite{Douglas:1997de}, and in fact trivialize the problem.

The equivalence between the dimer model and the formulation of \cite{Douglas:1997de} was proven in \cite{Franco:2006gc}, so instead of repeating the somewhat technical steps of this proof we instead take the pragmatic approach of showing how to construct the D3-brane linear sigma model from the dimer model, referring to \cite{Franco:2006gc} for the proof of equivalence to previous work.  This approach has the advantage of considerably simplifying the construction of the moduli space.

We saw in section \ref{s:diff-match} that the dimer configurations may be used to parametrize the D-term constraints of the gauge theory, via the contours that encircle the faces.  In fact the dimer variables also satisfy F-flatness.  Assign to each quiver field the product of matchings

\begin{equation}
 X_i = \prod_{\alpha=1}^{m} m_\alpha^{M_{i \alpha}}
\label{e:quivermap}
\end{equation}
i.e.~the product of all matchings that include the edge $X_i$ (here we have changed notation for the quiver fields: the subscript $i$ enumerates all fields).  Then the F-flatness constraints are of the form

\begin{equation}
 {\prod_{i \in V_1}}' X_i = {\prod_{j \in V_2}}' X_j
\label{e:f-prod}
\end{equation}
where the product is over the edges of two adjacent vertices $V_1, V_2$, and $\prod'$ indicates the product omits the edge connecting $V_1, V_2$; (\ref{e:f-prod}) is the F-flatness constraint associated to this omitted field.  By the map (\ref{e:quivermap}) and the definition of a perfect matching, each matching that appears on the LHS must also contribute to the RHS, and vice versa: thus all F-flatness constraints are trivially satisfied by (\ref{e:quivermap}) when written in terms of the matchings.

Recall that the matchings $m_\alpha$ are usually not linearly independent, and there are linear relations which may be represented by charge vectors $Q_F$.  We may then immediately use the results of section \ref{s:diff-match} to write down the linear sigma model for the classical moduli space of the D3-brane, by taking as charge vectors the face symmetries $Q_D^{a}$ (with associated FI parameters $\zeta^a$) and the redundancies $Q_F$ (which have no associated FI parameter since they do not represent gauge symmetries of the Lagrangian).

The reason that these additional ``fictitious'' $U(1)$ gauge symmetries appear in the linear sigma model is because we have used the perfect matchings to give a linear basis for the space of F-flat field configurations of the original theory, giving a toric subvariety of the space of D-flat vacua.  In other words, we have rewritten the F-terms as D-terms of a {\it different} field theory that has (by construction) the same classical moduli space.

What are the toric varieties described by this linear sigma model?  Recall that in general the charges $Q$ of a linear sigma model correspond to linear relations between the generators of the toric fan of the variety.  In fact, these are precisely the lattice generators specified by the Newton polygon of $\det K$, which we studied in section \ref{s:dimers-toric}.  To see this, recall from section \ref{s:diff-match} that the face symmetries (D-term constraints) are associated to linear combinations of matchings whose corresponding lattice vectors $(1,a,b)$ sum to $(0,0,0)$.  Similarly (by definition) the redundancies also correspond to identities between matchings, with total weight $(0,0,0)$.  In other words, the charges satisfy $Q . G^T = 0$ where $G$ is the matrix of $\Z^3$ lattice vectors of the Newton polygon, as required.  Thus, the moduli space of the quiver for the D3-brane on a Calabi-Yau manifold is the Calabi-Yau itself.  However, there is a subtlety with the identification of the moduli that we will discuss in more detail below.

\subsubsection{Example: Linear sigma model for classical del Pezzo 1 theory}

The linear sigma model for the moduli space of the classical gauge theory on the D3-branes at the conical del Pezzo 1 singularity is given by concatenating the charge matrices for the $N_F-1$ independent face symmetries $Q_D$, and the redundancies $Q_F$.  From section \ref{s:dp1-charges}

\begin{equation}
 Q=\left(\begin{array}{cccccccc|c}
		 0 & 0 & 0 & 0 & -1 & 0 & 0 & 1& \zeta^1\\
		 0 & 0 & 0 & 0 & 0 & 0 & 1& -1 & \zeta^2\\
		 0 & 0 & 0 & 0 & 0 & 1& -1 & 0 & \zeta^3\\
		 0 & 1 & 0 & 1 & 0 & -1 & 0 & -1 &0\\
		 1 & 1 & 1 & 0 & -1 & -1 & -1 & 0 &0\\
                        \end{array}\right)
\label{e:dp1-classical}
\end{equation}

The 8 matchings $m_i$ are charged under 5 gauge groups, giving a 3-dimensional space of vacua via the D-term equations

\begin{equation}
 \sum_{i=1}^8 Q_{i}^a |m_i|^2 = \zeta^a
\end{equation}
where $a=1, \ldots 5$ and $\zeta^4 = \zeta^5 = 0$.  We may easily confirm that $Q = ker G$, where

\begin{equation}
 G = \left( \begin{array}{cccccccc}
               m_1&m_2&m_3&m_4&m_5&m_6&m_7&m_8\\\hline
		1 & 1 & 1 & 1 & 1 & 1 & 1 & 1 \\
		-1& 0 & 1 & 0 & 0 & 0 & 0 & 0 \\
		0 & -1& 1 & 1 & 0 & 0 & 0 & 0 \\
            \end{array}\right)
\end{equation}
are the coordinates of the points in the toric diagram \ref{f:toric} for the Calabi-Yau cone over del Pezzo 1.   It is a straightforward exercise (see e.g.~\cite{Douglas:1997de,Beasley:1999uz}) to check that no matter what values the remaining $\zeta^{a = 1, \ldots, 3}$ take, the resulting space of vacua is always either the singular del Pezzo cone (if all $\zeta^a = 0$), a partial resolution of it (if some $\zeta^a \neq 0$), or a complete resolution to a smooth space (if $\zeta^a$ are generic).  In this case there is only one complete triangulation of the Newton polynomial so there is only a single smooth resolution for the cone, which corresponds to blowing up the vanishing del Pezzo cycle.  This triangulation as well as the dual web showing the finite-sized $dP_1$ are illustrated in figure \ref{f:toric}b.

By acting with $M^T$ we may recover the baryonic $U(1)$ charges of the quiver fields (as well as the two additional $U(1)$ that act trivially)

\begin{equation}
 Q.M^T = \left(\begin{array}{c|cccccccccc}
& X_{12} & X_{23}^1 & X_{23}^2 & X_{34}^1 & X_{34}^2 & X_{34}^3 & X_{42} & X_{13} & X_{41}^1 & X_{41}^2\\\hline
{\bf 1} & 1 & 0& 0& 0& 0& 0& 0& 1& -1& -1\\
{\bf 2} & -1 & 1& 1& 0& 0& 0& -1& 0& 0& 0\\
{\bf 3} & 0 & -1& -1& 1& 1& 1& 0& -1& 0& 0\\
&0 & 0& 0& 0& 0& 0& 0& 0& 0& 0\\
&0 & 0& 0& 0& 0& 0& 0& 0& 0& 0\\
\end{array}\right)
\end{equation}
which agrees with the quiver in figure \ref{f:dp1-quiver} via the incidence matrix (\ref{e:incidence}).  The charges under the gauge group {\bf 4} are fixed by the requirement that their sum over all gauge groups is zero.

\subsubsection{Comparison to Witten's 2d gauged linear sigma model}

The gauged linear sigma model we have constructed to study the moduli space of supersymmetric vacua of the D3-brane world-volume theory should not be confused with the gauged linear sigma model introduced by Witten to study topological properties of superstrings propagating on toric Calabi-Yau geometries \cite{Witten:1993yc}.  Witten's GLSM is a 2-dimensional gauge theory.  It is obtained by dimensional reduction from a 4-dimensional GLSM, and many of the essential properties it enjoys descend from the four-dimensional theory.  In fact, the linear sigma models constructed in this paper are of the four-dimensional variety.

The key distinction between the two-dimensional and four-dimensional theories is that the FI parameters in the latter are real, whereas in the Witten model each FI parameter pairs with the theta-angle associated to the $U(1)$ field strength, which is a scalar in 2 dimensions and is part of a twisted chiral superfield.  In 2 dimensions the FI parameter is complexified by the theta angle, but there is no analogous pairing in 4 dimensions because the FI term and theta-term do not combine into a superfield.

The four-dimensional nature of our linear sigma models arises since the $U(1)$ gauge groups are part of the $U(N)$ gauge groups of the four-dimensional D3-brane theory, which become global symmetries in the IR.  Therefore we may study the $U(1)$ subgroups in isolation and will later use them to classify the $SU(N)$ gauge-invariant operators of the CFT according to their $U(1)$ representations (charges).  The subtlety is that the variables of the linear sigma model are the dimer configurations (which automatically satisfy F-flatness), which are related to the quiver variables by a change of variables (as noted in the previous section we may also need to add additional $U(1)$ gauge factors that act trivially on all quiver fields).  We do not pursue the correspondence from the path integral point of view.

There are some other important differences from the Witten model.  Firstly, the D3-brane linear sigma model is {\it non-minimal}, in the sense that there are usually several fields with the same $U(1)$ charges (accordingly, there are additional $U(1)$ gauge groups in order that the moduli space remain 3-dimensional).  Finally, the FI parameters of the D3-brane linear sigma model are non-generic; some of them are fixed to zero.  In the Witten model they are both generic and complex, so the resulting phase structure is very different.

It is a highly nontrivial consequence of these three differences that no matter the choice of signs and values for the FI parameters of the D3-brane linear sigma model, the resulting vacuum space always has a geometrical description as a (possibly singular) Calabi-Yau variety, that is the $\zeta^a$ are always K\"ahler classes of the exceptional cycles of the Calabi-Yau geometry.  Varying the FI parameters may induce flops and other birational equivalences between Calabi-Yau spaces, but the ``non-geometrical'' phases of the Witten model for the superstring do not appear.  See \cite{Douglas:1997de,Beasley:1999uz} for further discussion of this issue.

It is also important to emphasize that the conical Calabi-Yau geometries we study correspond to {\it different} FI parameters for the Witten and D3-brane linear sigma models.  In the former the FI parameters are at the the orbifold point of the Calabi-Yau, which is typically located outside of the classical K\"ahler cone and is in a ``non-geometrical phase''.   In the latter the FI parameters are at the origin of the classical K\"ahler cone.  The Witten model for the superstring is actually singular at this point: this is because there are new light degrees of freedom corresponding to D-branes wrapped on a vanishing cycle of the Calabi-Yau that become massless.  The interpretation of this mismatch is that the D3-branes probe the underlying Calabi-Yau differently than do closed strings, and do not couple to the non-geometrical degrees of freedom of the closed string background \cite{Douglas:1997de}.

\subsection{Anomalous $U(1)$ symmetries}
\label{s:anom}

In order to study the IR CFT and its dual AdS model, we will need to account for the fact that some of the gauged $U(1) \subset U(N)$ symmetries of the UV theory are anomalous.  In general there will be two such anomalous $U(1)$'s \footnote{They are associated to the compact cycles of the toric surface at the base of the CY cone for which the dual cycle is also compact, namely the canonical class and the surface itself \cite{Buican:2006sn}.}.  In string theory the anomalies are cancelled by a generalized Green-Schwarz mechanism \cite{Douglas:1996sw,Ibanez:1998qp}, and the result is that the $U(1)$ gauge fields of the D3-brane world-volume theory couple to closed-string fields and become massive; they do not remain dynamical in the IR.  As emphasized in \cite{Intriligator:2005aw}, the D-terms of anomalous $U(1)$ factors should not be used to constrain the vacuum moduli space.

In the AdS model, the anomaly-free ``baryonic'' symmetries $U(1)$ remain gauged.  The corresponding $U(1)$ gauge field comes from the reduction of the RR 4-form on the 3-cycles of the Sasaki-Einstein space $X_5$.  These gauge fields couple to the global $U(1)$ currents of the CFT on the boundary of AdS.  The corresponding dibaryons of the CFT are dual to wrapped D3-branes, which are charged under these $U(1)$. As we noted in section \ref{s:toric}, the 3-cycles of a toric Sasaki-Einstein space are determined by the toric divisors of the CY cone, which may be easily read off from the linear sigma model.  This gives us a way to immediately identify the non-anomalous $U(1)$ symmetries of the CFT in terms of the dimer model.

Recall from \ref{s:dimers-toric} that the external matchings of the dimer model correspond to points on the boundary of the Newton polygon.  We used pairwise differences of these external matchings to define the ``zig-zag'' $U(1)$ symmetries of the CFT.  These symmetries fulfill the criteria of the previous paragraph and can be shown \cite{Butti:2005vn,Imamura:2006ie} to be precisely the anomaly-free $U(1)$ baryonic symmetries of the CFT.  The number of these zig-zag symmetries is given by $N_d-1$, where $N_d$ is the integer length of the boundary of $\Delta$.  As we discussed in section \ref{s:dimers-toric} the two primitive winding cycles are dual to mesonic symmetries, so the purely baryonic combinations of the zig-zag symmetries must have zero winding number.  This imposes two constraints on $N_d-1$ integer coefficients, so there are $N_d-3$ baryonic combinations of the zig-zag paths.  This is precisely the number of 3-cycles of $X_5$\footnote{There may be additional torsion cycles, for example Abelian orbifolds $S^5/\Z_n$ have a $Z_n$ torsion cycle that lifts to the trivial cycle in $S^5$.  Torsion cycles do not have a corresponding gauged $U(1)$, but there is a $\Z_n$ quantum symmetry that acts as $\Z_n$-valued baryon number \cite{Morrison:1998cs}.}.

Since the baryonic combinations of zig-zags have zero winding number on $T^2$, they must be equivalent to linear combinations of the face symmetries and redundancies (which are trivial in $H^1$).  This allows us to recover the relation to the $U(1) \subset U(N)$ ``classical'' baryonic symmetries and in particular to determine the assignment of FI parameters.

\subsubsection{Example: The anomaly-free baryonic symmetry of {del~Pezzo~1}}

We read off from figure \ref{f:dp1-normal} that the winding numbers of the 4 zig-zag symmetries for del Pezzo 1 are $(-1,-1), (2,-1), (0,1), (-1,1)$ respectively; only 3 of these are independent.  We look for a solution of

\begin{equation}
 a (-1,-1) + b(2,-1) + c(0,1) = (0,0)
\end{equation}
and choose $a=2, b=1, c=3$; other choices give a linear multiple of this charge assignment to the quiver fields.  Via the charge matrix (\ref{e:dp1-qz}) we see that the corresponding linear combination of the charges $Z_i$ is

\begin{equation}
2 Z_1 + Z_2 + 3 Z_3 =(2 , -1 , 2 , -3 , 0 , 0 , 0 , 0 | ?)
\label{e:dp1-af} 
\end{equation}
where the combination of FI parameters is yet to be determined.  To find it, we try to express (\ref{e:dp1-af}) as a linear combination of the classical symmetries (\ref{e:dp1-classical}).  We find

\begin{equation}
(2 , -1 , 2 , -3 , 0 , 0 , 0 , 0)= (-2 , 1 , -1 , -3 , 2) . Q
\label{e:dp1-af2}
\end{equation}
where Q is the matrix (\ref{e:dp1-classical}).  Therefore the FI parameter in (\ref{e:dp1-af}) is given by $\zeta = -2 \zeta^1 + \zeta^2 - \zeta^3$.  In particular, it may be either positive or negative.

The linear combination (\ref{e:dp1-af2}) is the only non-anomalous baryonic $U(1)$ of this theory; therefore the other anomalous baryonic symmetries $Q_D$ should not be treated as giving rise to D-term constraints.  That is, the $\zeta^a$ are not fixed parameters that constrain the allowed values of $|m_i|^2$; rather the $m_i$ charged under those $U(1)$ may take on any values, and $\zeta^a$ is {\it defined} to be equal to the corresponding linear combination of the $|m_i|^2$.

However, the redundancies $Q_F$ do still constrain the vevs of the fields.  In this example we may use the two redundancies to fix e.g.~$|m_8|^2 = |m_2|^2 + |m_4|^2 - |m_6|^2$ and $|m_7|^2 = |m_1|^2 + |m_2|^2 + |m_3|^2 - |m_5|^2 - |m_6|^2$, and we may then define

\begin{eqnarray}
 \zeta^1 &=& |m_8|^2 - |m_5|^2 =  |m_2|^2 + |m_4|^2 - |m_5|^2 - |m_6|^2\\\nonumber
 \zeta^2 &=& |m_7|^2 - |m_8|^2 =  |m_1|^2 + |m_3|^2 - |m_4|^2 - |m_5|^2\\\nonumber
 \zeta^3 &=& |m_6|^2 - |m_7|^2 = -|m_1|^2 - |m_2|^2 - |m_3|^2 + |m_5|^2 + 2 |m_6|^2
\end{eqnarray}
The non-anomalous combination of $\zeta^a$ is
\begin{equation}
 -2 \zeta^1 + \zeta^2 - \zeta^3 = 2 |m_1|^2 - |m_2|^2 + 2 |m_3|^2 - 3 |m_4|^2
\label{e:dp1-af3}
\end{equation}
i.e.~the non-anomalous baryonic symmetry corresponds to the sigma model defined over the {\it external} matchings only, with charges 
\begin{equation}
Q = \left( \begin{array}{cccc|c}
      2 & -1 & 2 & -3 & \zeta
      \end{array}\right)
\label{e:dp1-lsm}
\end{equation}
and the FI parameter $\zeta = -2 \zeta^1 + \zeta^2 - \zeta^3$.  Note that $\zeta$ {\it does not depend} on the internal matchings, since they entirely cancel from the expression (\ref{e:dp1-af3}).  $\zeta$ is a legitimate FI parameter for the anomaly-free gauged $U(1)$ baryonic symmetry and does constrain the vacua of the theory; therefore the vacuum geometry is described by the effective LSM (\ref{e:dp1-lsm}).

\begin{figure}[t]
 \begin{center}
  \epsfig{file=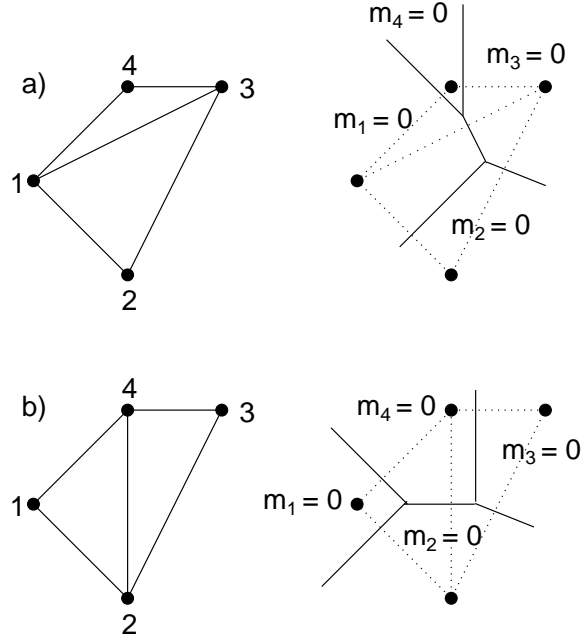,width=3in}
 \end{center}
\caption{The two possible partial desingularizations of the complex cone over del Pezzo 1 that are realized by the linear sigma model for the anomaly-free baryonic $U(1)$ symmetry.  a) The phase $\zeta > 0$.  By the charge matrix (\ref{e:dp1-lsm}) both $m_2$ and $m_4$ cannot simultaneously vanish, so the corresponding toric divisors $m_2=0$ and $m_4=0$ do not intersect and are separated by a blown-up $\P^1$.  b) The phase $\zeta < 0$.  Here the divisors $m_1=0$ and $m_3=0$ do not intersect.\label{f:dp1-partial}}
\end{figure}

When $\zeta = 0$ this is the singular cone over the del Pezzo 1 surface.  For $\zeta \ne 0$ the geometry is a partial resolution of this singularity; there are two partial resolutions corresponding to the two choices of sign for $\zeta$, giving the two possible partial triangulations of the toric diagram with the internal point removed, see Figure \ref{f:dp1-partial}.  In both resolutions $|\zeta|$ is the K\"ahler class of a $\P^1$; when $\zeta > 0$ there is a $\C^2/\Z_2$ singularity at every point on this $\P^1$, and when $\zeta < 0$ there is a point on the $\P^1$ with an isolated $\C^3/\Z_3$ singularity.  These facts may be read off from the $(p,q)$-web description of the geometry.

Finally from (\ref{e:dp1-af}) we read off the baryonic charge assignments of the quiver fields by multiplying with $M^T$,
\begin{equation}
 Q_B.M^T = \left( \begin{array}{cccccccccc}
X_{12} & X_{23}^1 & X_{23}^2 & X_{34}^1 & X_{34}^2 & X_{34}^3  & X_{42} & X_{13} & X_{41}^1 & X_{41}^2\\\hline
-3&2&2&-1&-1&-1&-1&-1&2&2
\label{e:qb}
\end{array}\right)
\end{equation}

\subsection{The BPS spectrum of the CFT from paths on the dimer lattice}
\label{s:spectrum}

Once we have identified the anomaly-free linear combinations of the gauged $U(1)$ symmetries, the corresponding FI parameters do not receive quantum corrections (in four dimensions they are only corrected at one loop, and this correction is proportional to $\sum_i Q_i$ and vanishes in the CY case; this fact was applied by Witten in studying the dimensional reduction to two  dimensions \cite{Witten:1993yc}).  $\N=1$ supersymmetry prevents corrections to the form of the superpotential in perturbation theory.  The effect of RG flow is that the gauge couplings of the anomaly-free $U(1)$ factors flow to zero and become global symmetries, and the massive $U(1)$'s corresponding to the anomalous symmetries become non-dynamical and decouple.  The BPS spectrum of the CFT in the IR is classified by the anomaly-free global $U(1)$ baryonic and mesonic charges of the operators, and these states may be matched to states of the linear sigma model.

In fact both the mesonic and baryonic BPS operators of the CFT have a natural representation in terms of paths on the brane tiling.  The former are represented by ``closed strings'' and the latter by ``open strings'' with fixed endpoints.  It is convenient to pass to the covering space of the $T^2$ in which the brane tiling is extended to a doubly-periodic tiling of the infinite plane.

\subsubsection{Mesonic operators}

A mesonic operator is formed from a closed path on the quiver, which is dual to a closed path on the faces of the brane tiling on $T^2$.  This gives a sequence $\{X_i\}$ of quiver operators whose product transforms in the adjoint representation of a gauge group; we then trace over these gauge indices to form the mesonic operator, $\Tr \prod_i X_i$.

Such an operator looks like a closed string with certain winding number (possibly vanishing) around the two cycles of the torus (recall the winding number records the global $U(1)^2$ mesonic charges of the operator).  Lifting to the covering space, it is a path from a given face to an image of the same face.  The classical BPS mesons correspond to strings of operators that are either holomorphic or antiholomorphic, i.e.~which only cross edges in a direction compatible with the bipartite orientation of the graph (e.g.~crossing an edge with the black vertex on the right)\footnote{In the quantum gauge theory one wishes to construct a basis of operators with definite conformal dimension, which requires diagonalizing the $n$-loop dilatation operator of the gauge theory.  In general operators will mix with one another order by order in perturbation theory, which makes constructing the eigenstates more difficult.  However, at one loop order the ``holomorphic'' (similarly anti-holomorphic) scalar BPS operators only mix among themselves and not with the other fields of the gauge theory \cite{Wang:2003cu} and one may define a spin chain model in terms of these same strings propagating on the brane tiling, where the symmetry group is the $U(1)^3$ global flavour group of the CFT (or its non-Abelian extension), and the spin chain interactions come from the F-term relations of the quiver theory.  One may conjecture that this model is integrable for a general toric theory; this is presently under investigation \cite{KennawayWIP}.}.

The $U(1)$ charges of this operator are given by summing the charges of the edges of the tiling crossed by the closed string.  Mesonic operators have zero baryon number and map via $M$ to linear combinations of the matchings of the linear sigma model with vanishing GLSM charge.  Thus, they are global holomorphic sections of the $U(1)^B$ gauge bundle of the LSM, and correspond to $U(1)^B$ gauge-invariant combinations of the matchings.  These global holomorphic sections give a natural lattice structure to the real cone spanned by the D-term equations of the LSM \cite{Hanany:2006nm,Benvenuti:2006qr,Martelli:2006vh}.

\begin{figure}[t]
 \begin{center}
  \epsfig{file=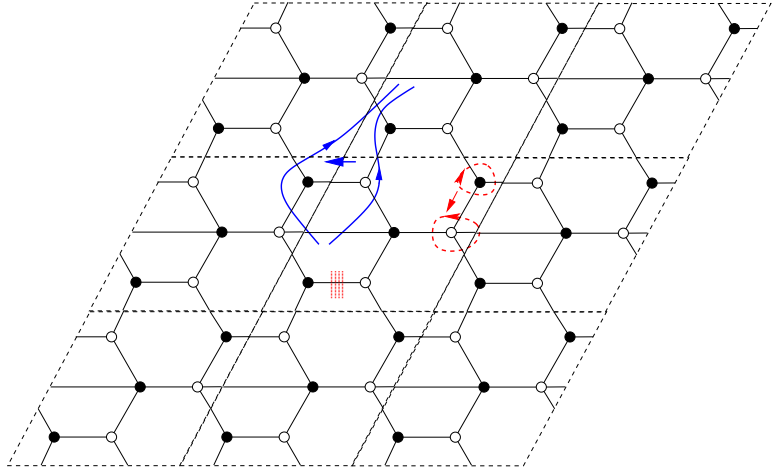}
 \end{center}
\caption{A $3 \times 3$ subset of the infinite cover of the brane tiling for del Pezzo 1 showing some gauge-invariant operators as open and closed strings.  We choose the center tile as the origin.  Clockwise from top left: a) the solid lines represent mesons; they are closed strings stretching from a face to another image of that face after winding around the torus (the winding number $(0,1)$ of this closed string is accounted for by the lift to the infinite cover).  We have shown two F-term equivalent mesonic operators, $\Tr X_{12} X_{23}^1 X_{34}^1 X_{41}^1 \sim \Tr X_{12} X_{23}^2 X_{34}^2 X_{41}^1$.  b) Two F-term equivalent mesonic operators with zero winding number, which encircle vertices.  c) A baryonic operator $\det(X_{13})$. \label{f:dp1-cover}}
\end{figure}

In general there are many mesonic operators mapping to a given gauge-invariant combination of the matchings.  However, recall that the role of the matchings is to trivialize the F-term relations of the quiver theory.  Thus, this many-to-1 map from mesons to matchings becomes 1-1 after we impose F-term relations on the mesons.  In terms of the brane tiling this equivalence has a nice interpretation.  The F-term relations may be used to move the ``closed string'' mesonic operator from one configuration to an equivalent configuration, which in particular has the same R-charge.  All configurations with the same winding number and R-charge are F-term equivalent; thus the closed string may propagate on the torus, with constant ``string length'' (which we may define to be equal to its R-charge; the number of fields in the operator may however fluctuate), see figure \ref{f:dp1-cover}.  The enumeration of mesonic operators is translated to a counting problem of closed paths on the infinite cover of the brane tiling, from a fixed reference face to images of that tile.  This gives a lattice structure to the infinite tiling, which in fact agrees with the lattice of gauge-invariant monomials on the linear sigma model \cite{Benvenuti:2006qr,Hanany:2006nm}.

It is possible to show \cite{Hanany:2006nm} that for the toric quiver theories the gauge-invariant operators are completely classified by their $U(1)$ charges.  In the present context all holomorphic paths between two given endpoints, with the same $U(1)_R$ charge, are F-term equivalent.  This has an important consequence: if we consider a general mesonic operator formed by composing several ``primitive'' mesonic building blocks (those paths which come back to an image of their starting gauge group exactly once),

\begin{equation}
\Tr M_1 M_2 \ldots M_n 
\end{equation}
then according to the result of \cite{Hanany:2006nm} the ordering of the $M_i$ is irrelevant up to F-term relations.  In other words, all primitive mesonic building blocks $M_i$ commute inside the trace, and may be simultaneously diagonalized.  Thus, the eigenvalues of the primitive mesons are coordinates on the moduli space of mesonic vacua, up to permutation.  We will discuss this further in the next section.

\subsubsection{Baryonic operators}

Baryonic operators transform with non-zero charge under the anomaly-free $U(1)^B$ symmetries.  Since these are gauge symmetries in AdS or the linear sigma model, the operators correspond to holomorphic sections of a $U(1)^B$ line bundle over the space of D-term vacua.  What is this line bundle?  The answer again comes from the matching matrix $M$.  Recall from (\ref{e:quivermap}) that the map from quiver fields to matchings is given by:

\begin{equation}
 X_i = \prod_{\alpha=1}^{m} m_\alpha^{M_{i \alpha}}
\label{e:quivermap2}
\end{equation}
As we saw in section \ref{s:anom}, it is only the external matchings that generate the anomaly-free baryonic symmetries, therefore we should restrict the sum to the external matchings (a subset of the columns of $M$).  As we discussed in section \ref{s:dimers-toric}, the external matchings also generate the toric divisors $D_i = \{ m_i = 0\}$, which correspond to 3-cycles of the Sasaki-Einstein space $X^5$.  The map from quiver fields to line bundles is given by (up to linear equivalence of divisors) \cite{Hanany:2006nm}

\begin{equation}
 X_i \mapsto {\cal O}(\sum_{\alpha=1}^{m_{ext}} M_{i \alpha} D_\alpha)
\end{equation}
In general the bundles ${\cal O}(\sum a_i D_i)$ may have more than one holomorphic section; the multiplicity of baryons associated to each divisor may be easily computed using toric geometry methods \cite{Franco:2005sm}.

Turning on a non-zero FI parameter in the AdS model forces some of the baryonic operators to develop vacuum expectation values.  The FI parameter breaks conformal invariance by introducing a scale, and the theory will undergo RG flow.  The new IR theory will be described by the corresponding partial resolution of the singularity as prescribed by the linear sigma model; a complete resolution gives a smooth Calabi-Yau which has an accidental $\N=4$ supersymmetry in the IR.

In the CFT the $U(1)$ are global symmetries, so the baryonic operators are states of definite charge, and descend from the holomorphic sections of the $U(1)$ bundles.  Turning on the FI parameter is a relevant deformation of the CFT.  Nevertheless, there is a discrete remnant of the FI parameter that classifies the spectrum: it was shown in examples in \cite{Forcella:2007wk} that the problem of counting operators with definite baryon number $\mu$ (i.e.~sections of a nontrivial line-bundle on the space of D-term constraints with all FI parameters $\zeta=0$) is equivalent to counting ${\it global}$ sections, i.e.~functions, on the resolution of this space, with FI parameter $\zeta=\mu$.  In other words, the $\Z$-valued baryon numbers of the CFT can be identified with the FI parameters of the linear sigma model, and in particular are quantized.

From the point of view of the brane tilings, we may note that the baryonic operators again have a simple representation on the infinite cover of the brane tiling.  The basic building blocks are an ``open string'' of operators, whose two endpoints each have a free gauge index.  Taking $N$ copies of this operator and contracting the free indices with two epsilon tensors produces a dibaryonic operator as in (\ref{e:baryon}).

As with the mesonic operators, we are free to move the interior of this collection of $N$ open strings around on the lattice by applying F-term relations, but the endpoints are fixed to lie on the two given faces of the tiling (fractional D-branes).  Thus we again reduce to a problem of enumerating possible endpoints of paths in the infinite cover of the dimer lattice.

Presumably one could understand the chiral ring relations of the CFT in terms of splitting and joining of these open and closed strings; it would be interesting to develop this correspondence further and to relate it to the dynamics of the Type IIB string theory (or perhaps the Type IIA theory, see section \ref{s:geom}).

\subsection{The BPS vacua}

When all FI parameters are set to zero, the space of classical LSM vacua is isomorphic to the toric CY cone, and this space is closely related to the mesonic vacua of the gauge theory.  The mesonic vevs are interpreted as the positions of the $N$ D-branes on the toric CY cone.  Since the mesonic operators commute inside the trace they may be simultaneously diagonalized, and the space of classical vacua is parametrized by the $N$ eigenvalues, which are coordinates on the space of LSM vacua.  Therefore the mesonic vacua are specified by choosing $N$ points on the toric CY cone $\cal M$ up to permutation, or equivalently one point in the space $Sym^N(\cal M)$.

A generic point in $Sym^N(\cal M)$ is smooth, and corresponds to $N$ separated D-branes on $\cal M$.  The world-volume gauge theory is broken to $U(1)^{N-1}$ coupled to adjoint matter.  This theory is IR free.

Singular points of $Sym^N(\cal M)$ correspond to coincident D-branes and enhanced world-volume symmetry.  When $n$ coincident D-branes are located at a smooth point of the space, the low energy theory is $\N=4$ $SU(n)$ Yang-Mills.  When they are located at a singular point (e.g.~the tip of the cone or its partial resolution), the low energy theory is the $SU(n)$ version of the corresponding $\N=1$ superconformal quiver theory.  In some cases the mesonic moduli space may also admit other branches corresponding to non-isolated singularities \cite{Morrison:1998cs}.
 
Turning on baryonic vevs changes the LSM parameters $\zeta^a$, which resolves the cone.  They correspond to deformations of the closed string background that blow up a cycle and introduce a mass scale.  At energies much less than this scale the massive fields become non-dynamical and may be integrated out.  This is equivalent to taking a suitable limit $\zeta^a \rightarrow \infty$ in which the size of the cycle, and the corresponding mass scale, goes to infinity.  In the infrared the theory is again conformal but describes D-branes propagating on a ``simpler'' singularity obtained by partially resolving the original one.  Note that after taking into account the anomalous $U(1)$ symmetries not all partial resolutions of the singularity may be accessible, as in the del Pezzo 1 example.

\subsection{Identifying the $U(1)_R$ charge: isoradial embeddings and $a$-maximization}
\label{s:iso}

An important problem that we have so far ignored is to study the $U(1)$ R-symmetry of the CFT.  We have also ignored the physical meaning of the edge weights appearing in the Kasteleyn matrix.  These turn out to be closely related.  Recall that we used the vanishing $\beta$-function constraint (\ref{e:scft-v}) and the requirement that W has R-charge 2 (\ref{e:r-charge}) to determine that the brane tiling is embedded into a 2-torus.   In fact we can do better and determine a family of embeddings into a {\it flat} torus.

Pick a point on the interior of each face, and for each edge $i$ of the graph consider the angle $\alpha_i$ subtended between the two interior points that includes this edge.  If we assign to the corresponding quiver field $X_i$ the R-charge

\begin{equation}
R(X_i) = \frac{\alpha_i}{\pi}
\label{e:norm-r}
\end{equation}
then equation (\ref{e:scft-v}) is automatically satisfied by this assignment, since the sum of the angles around the vertex is $2 \pi$ (thus, the superpotential term has R-charge 2).  In fact, since each edge connects two vertices, (\ref{e:norm-r}) imposes a constraint that the angles subtended from both vertices must be equal, and the quadrilateral formed by connecting the two interior points to the vertices of the brane tiling must be a diamond.  This is a constraint on the allowed embedding of the brane tiling.

\begin{figure}[th]
\begin{center}
 \epsfig{file=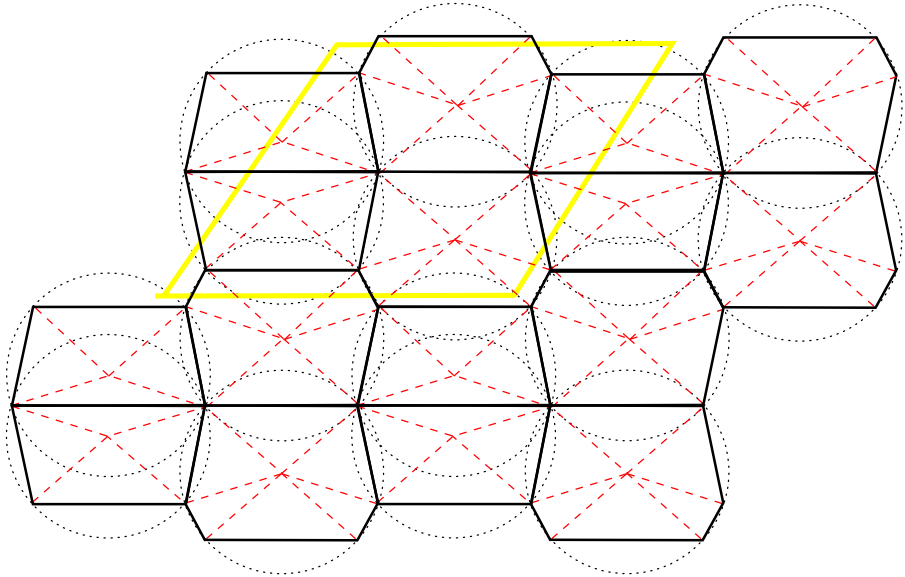,height=3in}
\end{center}
 \caption{An isoradial embedding of the del Pezzo 1 tiling, showing several fundamental domains\label{f:dp1-iso}}
\end{figure}

In fact we can say more: the vanishing of the $\beta$-function (\ref{e:scft-v}) for a gauge group says that

\begin{equation}
 \sum_{i \in \mbox{face}} (\pi - \alpha_i) = 2 \pi
\end{equation}
If we require that $(\pi - \alpha_i)$ is the other angle of the diamond, i.e.~the angle subtended by the edge from the interior point of the face, then the diamond is actually a {\it rhombus}.  In particular, all four edges of a rhombus have equal length, and the vertices of the tiling must all be equidistant from the interior points.  In other words, we must choose an embedding of the graph such that the polygonal boundary of the face of the tiling is inscribed in a circle, with our chosen interior point at the center.  The same argument applies equally to all faces of the tiling, and we have restricted the allowed embeddings of the tiling to the class of {\it isoradial embeddings}, those where every face of the tiling is inscribed in a circle of equal (e.g.~unit) radius.  See figure \ref{f:dp1-iso} for an example of an isoradial embedding of the del Pezzo 1 tiling.


\begin{figure}[th]
\begin{center}
 \epsfig{file=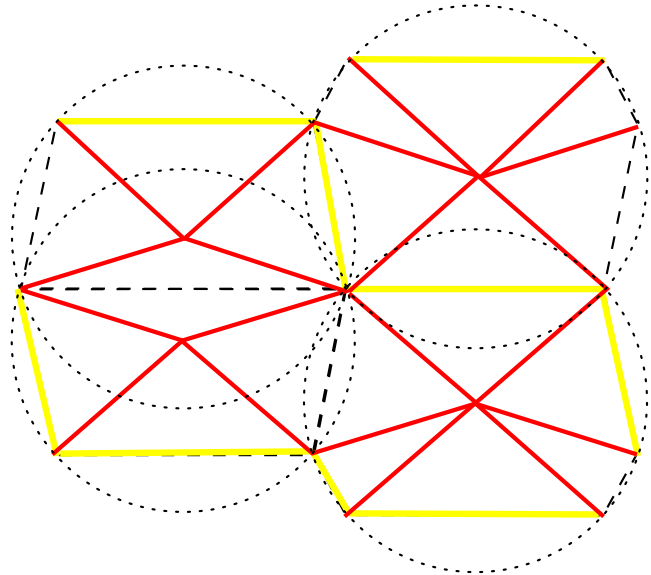,height=3in}
\end{center}
 \caption{Zig-zag paths define rhombus chains, sequences of rhombi whose opposite edges are all parallel along the chain.\label{f:dp1-iso-zz}}
\end{figure}

Now the zig-zag paths also play a special role in the isoradial embeddings.  Observe from figure \ref{f:dp1-iso-zz} that the zig-zag paths define a sequence of rhombi which share a common angle; i.e.~one pair of opposite sides of the rhombi in the sequence are all parallel.  We are free to deform the angles $\alpha$ along such a zigzag path by alternately ``stretching'' and ``squeezing'' successive rhombi by a common angle $+\delta, -\delta$.  Recall that the zigzag path is the difference of two matchings, so the charges of the quiver fields under the associated $U(1)$ alternate $+1, -1$.  So this operation is precisely a mixing of the $U(1)_R$ symmetry with the $U(1)$ zigzag symmetry, which is a linear combination of the anomaly-free $U(1)$ symmetries of the CFT.  We may mix with all of the zig-zag symmetries in this way, subject to the constraint that the R-charge of the fields lies in the interval $[0,1]$ (the endpoints correspond to a degenerate rhombus with angles $0, \pi$)\footnote{In fact, some tilings cannot be isoradially embedded unless some of the angles are degenerate.  It is conjectured that one may always perform a Seiberg duality to obtain a non-degenerate tiling.}.

According to the proposal of \cite{Intriligator:2003jj}, the true R-charge assignment of the IR fixed point is given by maximizing the central charge $a$ of the theory as a linear function of the various global $U(1)$ charges of the fields.   The $a$-function is given by

\begin{equation}
 a(R) = \frac{3}{32}(3 \Tr R^3 - Tr R)
\end{equation}
where the trace is over all states.  For the quiver theories anomaly cancellation implies $\Tr R = 0$, and the trace becomes

\begin{eqnarray}
 a &=& \frac{9}{32} \sum_i (R_i - 1)^3\nonumber\\
&=& -\frac{9}{32 \pi^3} \sum_i \beta_i^3
\label{e:amax}
\end{eqnarray}
where $\beta_i \equiv \pi - \alpha_i$ is the other angle of the rhombus whose diagonal contains the field $i$.

Thus, the procedure of $a$-maximization that determines the true R-symmetry of the CFT may be represented as an extremization of the cubic function (\ref{e:amax}) over the moduli space of isoradial embeddings of the brane tiling, which is the space of angles $\beta_i$.  $a$-maximization was recast as an operation in toric geometry (``Z-minimization'') in \cite{Martelli:2005tp}, and the two prescriptions were proven to be equivalent in \cite{Butti:2005vn}.  A mirror version of Z-minimization will be briefly discussed in section \ref{s:spectral}.

Given an isoradial embedding, it is natural to equate the edge weights appearing in the Kasteleyn matrix to the distance between the centers of the two circles adjacent to the edge \cite{Kenyon:2003ui}, i.e.

\begin{equation}
 e_i = \pm 2 \cos(\frac{\pi - \alpha_i}{2}) = \pm 2 \cos(\frac{\pi}{2}(1-R(X_i)) 
\label{e:weights}
\end{equation}
Note that the edge weights are valued between 0 and 2 in absolute value and we have allowed for the possibility of a sign.  We will return to this, and discuss the role of this sign choice, in section \ref{s:spectral}.  The isoradial embeddings will turn out to be related to the mirror Calabi-Yau geometry.
\section{Geometry of brane tilings from mirror symmetry}
\label{s:geom}

Our goal now is to show how the brane tilings appear in string theory.  We will first revise some relevant background of mirror symmetry.  We will show that the data of the brane tilings already encodes the structure of the mirror geometry, and in fact some applications of techniques from graph theory and knot theory will provide the connection between the brane tilings as we have previously studied them, and the geometry of (special) Lagrangian 3-cycles in the mirror Calabi-Yau geometry.

\subsection{Hori-Vafa vs geometrical mirror symmetry}
\label{s:hv}

The Hori-Vafa construction \cite{Hori:2000ck} gives a procedure for associating a 2-dimensional ``mirror'' QFT to a Witten linear sigma model (possibly with superpotential, if we wish to describe string theory on a compact manifold).  The claim is that on both sides of the duality the IR dynamics of the QFT has a nontrivial conformal fixed point, and recovers the propagation of the superstring on a CY and its mirror (or more general situations).  Explicitly, for a Witten linear sigma model defined by a set of charge vectors $Q^a$ and associated (complexified) FI parameters $t^a$ (with no superpotential), the Hori-Vafa mirror is a Landau-Ginzburg model with superpotential given by

\begin{equation}
 W_{HV} = \sum_i e^{-Y_i}
\label{e:WHV}
\end{equation}
where the fields $Y_i$ are dual to the LSM fields $X_i$, and are related via

\begin{equation}
 Re Y_i = |X_i|^2
\end{equation}
Imposing holomorphy of $W$, the complex fields $Y_i$ are constrained to obey the analogue of the D-term constraints

\begin{equation}
 \sum_i Q_i^a Y_i = t^a
\end{equation}
for each $U(1)$ gauge group $a$, so in general $a$ fields may be eliminated from the sum (\ref{e:WHV}) by imposing these constraints.  We expect that the mirror CFT should be described by the IR fixed point of this theory.  However in the Calabi-Yau case there is a difficulty because the solutions to the F-term equations $\frac{\partial W_{HV}}{\partial Y_i} = 0$ do not exist at finite distance in field space.  Indeed, by a linear change of variables $W_{HV}$ can always be brought to the form

\begin{equation}
 W_{HV} = e^{-x} P(e^{-y}, e^{-z})
\label{e:hv}
\end{equation}
where $x, y, z$ are suitable linear combinations of the $Y_i$, and $P$ is the Newton polynomial of the toric Calabi-Yau, which we have already encountered.  It is clear that this superpotential has a runaway behaviour along the positive $x$ direction, and there is no critical point at finite distance.

Nevertheless, it can be shown \cite{Hori:2000kt} that the periods of $W_{HV}$ (which measure the mass of BPS D-branes), agree with the periods of the manifold

\begin{equation}
 P(x,y; t) + u v = 0
\label{e:cy}
\end{equation}
where $x,w \in \C^*$, $u, v \in \C$ and $P(x,y; t)$ is the same Newton polynomial that appeared above; here we are emphasizing the complex structure moduli $t$ (coefficients of $P$).  This hypersurface in $(\C^*)^2 \times \C^2$ is a Calabi-Yau 3-fold.
 
Apart from the matching of periods, it is unknown how precisely to relate the Hori-Vafa mirror Landau-Ginzburg theory to the sigma model on the mirror Calabi-Yau target space.  Nevertheless, since this Calabi-Yau has the correct properties to agree with the mirror to the toric Calabi-Yau we started with, we will refer to it as the geometrical mirror.  In particular, since we are studying the properties of BPS D-branes, passing to the geometrical mirror is justified.


\subsection{The mirror D6-branes}

The SYZ conjecture \cite{Strominger:1996it} states that the mirror to a D3-brane at a point on a Calabi-Yau 3-fold should be a D6-brane wrapping a 3-torus in the mirror Calabi-Yau (mirror symmetry is roughly ``fibre-wise T-duality on a $T^3$ fibre'').  This should continue to hold true for a D3-brane at a singular point of the Calabi-Yau; the mirror to the singular point of the Calabi-Yau is a singular $T^3$.

As we have discussed, when the Calabi-Yau is located at the orbifold point in K\"ahler moduli space, a D3-brane at the singularity is better described by a collection of mutually supersymmetric fractional branes that are localized to the singularity (they are coherent sheaves supported on the vanishing cycle).  The fractional D-branes sum to give the D3-brane.
Under mirror symmetry these fractional branes map to D6-branes wrapping special Lagrangian 3-cycles of the mirror Calabi-Yau; these mirror fractional D6-branes should sum to the homology class of a singular 3-torus.  We will see how this arises later on.

The 3-cycles wrapped by the mirror fractional D6-branes may be constructed geometrically using techniques from singularity theory \cite{Hori:2000ck}.  The physical picture we expect is the following: each of the fractional D-branes produces a gauge group of the quiver theory, and their pairwise intersection supports a massless open string at the intersection point that gives rise to a bifundamental chiral multiplet in the four-dimensional gauge theory \cite{Berkooz:1996km}.  Superpotential terms come from worldsheet disc instantons; we will return to this later.

We begin by rewriting the defining equation (\ref{e:cy}) of the mirror Calabi-Yau, introducing a new $\C$-variable

\begin{eqnarray}
 W &=& P(w,z; t)\label{e:fib1}\\
 W &=& - uv\label{e:fib2}
\end{eqnarray}
which exhibits the Calabi-Yau as a double fibration over the complex $W$-plane.  Equation (\ref{e:fib1}) defines a holomorphic curve in $(\C^*)^2$ over every point in $W$, and (\ref{e:fib2}) a complex quadric.

Note that (\ref{e:fib2}) admits a $\C^*$ action by rescaling $(u,v) \sim (\lambda u, \lambda^{-1} v)$, and in particular a $U(1)$ action by phase rotations.  Since $u, v \in \C$, this $\C^*$ action degenerates above $W=0$, and the fibre contains a vanishing $S^1$.  Similarly, at certain points $W = W_*$, the holomorphic curve (\ref{e:fib1}) becomes singular,

\begin{equation}
 \frac{\partial (P(w,z; t) - W_*)}{\partial w} = \frac{\partial (P(w,z; t)-W_*)}{\partial z} = P(w,z; t) - W_* = 0
\end{equation}
this occurs when a 1-cycle of the curve pinches off, i.e.~the curve develops a vanishing cycle.  Generically these critical points $W_*$ are distinct from one another and from the origin, but their location varies as a function of the complex structure moduli $t$ of $P(w,z; t)$ and they may coincide at special loci in the moduli space (such as the conifold point, when one of the critical points approaches the origin and the wrapped D6-brane becomes massless).  For now we assume that the critical points are distinct.

For each critical point $W_*$, construct a path in the $W$-plane connecting $0$ and $W_*$.  The total space over this path is topologically a 3-sphere; this may be easily seen by considering the example of

\begin{equation}
 |z_1|^2 + |z_2|^2 = 1
\end{equation}
where $z_i = x_i + i y_i \in \C$.  This is the equation of a unit 3-sphere in $\C^2 \simeq \R^4$.  Projecting to the $|z_i|^2$ variables, it describes an interval; the fibre over the generic point on the interval is a $T^2$ corresponding to the angular parts of $z_i$, and over the two endpoints of the interval one or the other of the circles shrinks.  This is the same situation we have constructed in the mirror geometry, so we obtain a collection of 3-spheres corresponding to the straight line paths connecting the origin to the critical values in the W-plane.  In fact they can be chosen to be Lagrangian with respect to the natural symplectic form on $(\C^*)^2 \times \C^2$ \cite{Hori:2000ck}.  We expect (but do not prove) that at the orbifold point the straight line paths admit representative 3-cycles that are furthermore special Lagrangian with equal phase, corresponding to the mutually BPS fractional branes.

When the toric diagram contains one or more internal points, the toric Calabi-Yau admits a compact 4-cycle; similarly the curve $P(w,z; t)=0$ has genus $g \ge 1$.  Then the number of critical points of $P(w,z; t)$ (and hence the number of 3-spheres) is given by twice the area of the toric diagram.  This is the same as the number of triangles in any complete triangulation of the toric diagram.  This fact follows from the Bernstein-Koushnirenko theorem \cite{Feng:2005gw,Bernstein,Koushnirenko}.

Interestingly, the equality fails in the case when there are no compact 4-cycles, i.e.~the toric diagram does not contain an internal point, and the curve $P(w,z; t)=0$ is genus zero.  In this case the function $P(w,z; t)$ does not admit enough critical points to account for the fractional branes.  In this case the construction of the 3-cycles appears to be more subtle; nevertheless the methods we will introduce later continue to provide a formal construction of the correct brane tiling in terms of the curve $P(w,z; t)=0$.

By construction, these 3-spheres can only intersect in the fibre above $W=0$.  Their image in the quadric fibre (\ref{e:fib2}) above $W=0$ is always the single point at the origin in $\C^2$, so the question of their intersection reduces to the intersection in the other fibre, the holomorphic curve defined by

\begin{figure}[tb]
\begin{center}
 \epsfig{file=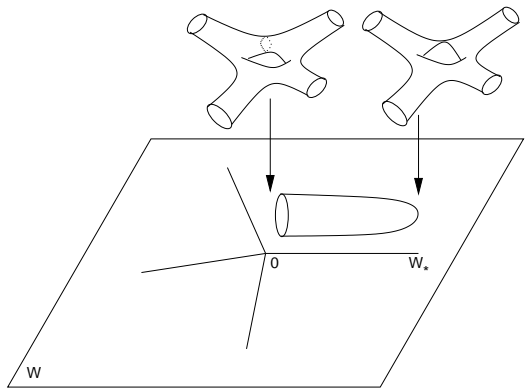}
\end{center}
\caption{The vanishing cycles of $P(w,z; t) = W$ at $W=W_*$ evolve Lagrangian disks in $(\C^*)^2$ along the path from $W_*$ to $0$.  The remaining $S^1$ fibre over the disk vanishes at $W=0$, i.e.~on the boundary of the disk, and gives the 3-cycle the topology of $S^3$.  These 3-cycles are $1:1$ with the gauge groups of the quiver theory and are wrapped by the fractional D6-branes.\label{f:disks}}
\end{figure}

\begin{equation}
\Sigma = \{P(w,z; t) = 0\} \subset (\C^*)^2
\end{equation}
In $(\C^*)^2$ the 3-cycles project to disks, see figure \ref{f:disks}.  Above $W_*$ there is a fixed point, and along the interval $[0, W_*]$, the fixed point is resolved into a 1-cycle of the curve $P(w,z; t) = W_*$ and sweeps out a disc. The Lagrangian 3-cycle projects to a Lagrangian disk bounded by a suitable representative of this vanishing cycle in the smooth curve $P(w,z; t) =0$; it is suspended from this boundary on $\Sigma \subset (\C^*)^2$ like a soap bubble.

The $S^1$ in the quadric fibre is fibered over this disk such that it vanishes on the boundary and is finite in the interior.  These disks do not intersect one another except on their boundary.  This follows because every point $(w_0, z_0) \in (\C^*)^2$ lies on the curve fibre above precisely one point: this point $W_0$ is given by evaluating $W_0 := P(w_0,z_0; t)$ (furthermore we also see that the family of curves $P(w,z; t) = W$ foliates the space $(\C^*)^2$, so we may consider this $(\C^*)^2$ as a subspace of the Calabi-Yau 3-fold).  Thus, the intersection problem is reduced to determining the intersection on the curve $\Sigma$ of the 1-cycles that are the boundaries of the disks.

In simple cases \cite{Hanany:2001py,Orlov} one may compute the intersection number of the 1-cycles on the $\Sigma$, which measures the net number of (signed) intersections counted with their orientations.  In some cases this is enough to reconstruct the quiver, but it misses non-chiral matter (bidirectional arrows in the quiver, which cancel out from the sum).  In the general situation (when the base of the toric CY cone is itself not smooth) the quiver contains such non-chiral matter.  

In the general situation the intersection theory of these 1-cycles has not been directly computed.  However, we will use a different route to construct a consistent set of intersecting 1-cycles, which precisely reproduce the data of the quiver including the non-chiral matter.  We conjecture that this indeed produces the correct Lagrangian vanishing cycles, but this has not been proven in general.  We will then give some supporting evidence for this conjecture based on a closer analysis of the geometry.  The construction proceeds by a closer analysis of the graph theory underlying the brane tilings.

\subsection{The spectral curve and the mirror D6-brane moduli space}
\label{s:spectral}

As a first step, we show that the combinatorics of dimer models encodes the essential part of the mirror Calabi-Yau geometry, the holomorphic curve $P(w,z) = 0$, as well as (with an additional assumption) the expected view of the moduli space as probed by the mirror D6-branes.

Recall that the key property of the Kasteleyn matrix $K(w,z)$ is that its determinant enumerates the dimer configurations of the brane tiling, which define the global symmetries of the CFT.  We may also use the Kasteleyn matrix $K(w,z)$ to define a holomorphic curve in $(\C^*)^2$, called the {\it spectral curve} of the dimer model, via

\begin{equation}
 \det K(z,w) = 0 \label{e:spectral}
\end{equation}
This is the same family of holomorphic curves that is used to construct the geometrical mirror Calabi-Yau manifold, studied in the previous section.  However, the values of the complex structure moduli (coefficients of the curve) differ.  With the isoradial rule (\ref{e:weights}) for assigning the edge weights (coefficients in $K$), the coefficients of (\ref{e:spectral}) are real (and the curve admits a real involution by complex conjugation).

In fact, with an additional choice of signs for the entries of the matrix $K(z,w)$, this real curve always has genus 0 \cite{Kenyon:2003ui}.  Moreover, it is a {\it Harnack curve}, meaning it has a maximal number of ovals (the genus 0 curve has all of the compact ovals degenerate, i.e.~they correspond to vanishing cycles).  We are instructed to multiply each edge weight $e_i$ by a sign $\pm 1$ such that the product of the edges bounding a face has parity given by

\begin{equation}
\prod_{e_i \in F_a} \mbox{sign}(e_i) = \left\{ \begin{array}{cc}
                             -1 & \quad\mbox{if } |F_a| = 0 \mod 4\\
                             +1 & \quad\mbox{if } |F_a| = 2 \mod 4
                            \end{array}\right.
\end{equation}
where $|F_a|$ is the number of sides of the face $F_a$.  It is always possible to achieve this \cite{Kasteleyn}, and the combinatorial effect is that all of the summands of the coefficients of $z^a w^b$ in the expansion of $\det K(w,z)$ have the same sign.

The physical justification for this sign rule is still mysterious.  Physically it can be accounted for by introducing a half-integral B-field through one or more of the vanishing cycles of the singularity, which shifts $e^t = e^{J + i B} = e^{J+i \pi} = - e^{J}$.  The origins of this shift deserve further study; however, the consequence of introducing this shift is clear.  Unless the sign rule is imposed, the spectral curve of the isoradially embedded brane tiling is not genus 0.  With the sign choice, it is precisely at the singular ``conifold'' point in complex structure moduli space, i.e.~the Calabi-Yau

\begin{equation}
 \det K(w,z; \bar t) + uv = 0 \label{e:notmirror}
\end{equation}
is singular, and we have emphasized the complex structure moduli $\bar t$ that are mirror to the K\"ahler moduli of the geometry seen by the D3-branes.  This agrees with the physical expectation that the geometry as seen by the D6-branes should be mirror to the geometry as seen by the D3-branes.

Thus, we have agreement with the proposal (\ref{e:weights}) relating the weights of the Kasteleyn matrix to the edge lengths of the embedding, and in turn to the role of angles as R-charges.  One needs to keep in mind that (just as in the D3-brane case), the complex structure moduli $\bar t$ as seen by the mirror D6-branes are not equal to the moduli of the closed string background, so (\ref{e:notmirror}) is not equal to the equation (\ref{e:cy}) of the ambient Calabi-Yau as probed by closed strings.

Recalling the discussion from section \ref{s:iso}, it was mentioned that the physical procedure of $a$-maximization has an equivalent presentation in toric geometry in terms of ``Z-minimization'', the minimization of a certain functional over a 2-dimensional parameter space, which corresponds physically to mixing with the two mesonic $U(1)$ symmetries of the CFT (the bosonic symmetries in fact decouple from $a$-maximization \cite{Butti:2005vn}).  We remark that there is a mirror version of this procedure: rescaling the coordinates $z, w$ provide coordinates on the moduli space of genus 0 Harnack curves, which are $1:1$ with the isoradial embeddings \cite{Kenyon:2003ui}.  As we saw in section \ref{s:quiver-dimer}, these coordinates are indeed associated to the two winding cycles of the torus, and to the mesonic $U(1)$ symmetries.  The mirror role of the $Z$ function is not currently understood.

\subsection{Some graph theory}
\label{s:graph}

We now turn to a deeper study of the graph theoretical properties of the brane tilings.  We will formalise some of the relevant properties of the brane tilings that we have used in previous sections.  This will lead us to a deeper understanding of the relationship between the brane tilings on $T^2$ and the geometry of the mirror D6-branes that engineer the quiver gauge theory.

\subsubsection{Zig-zag paths}

Given an embedding of a graph into an oriented Riemann surface, at every vertex there is an ordering of the incident edges induced by the orientation, given by proceeding e.g.~clockwise around the vertex from a given edge.  For each of the two possible orientations of an edge of the graph, we may define two {\it zig-zag paths} passing through this edge (up to overall orientation), which are oriented paths on the graph that alternately turn maximally left, then maximally right, at each successive vertex.  If a zig-zag path is closed (comes back to its starting point), then it is called a {\it zig-zag cycle}.  We already encountered examples of zig-zag cycles in section \ref{s:basis}, although we constructed them somewhat differently.

Note that the zig-zag paths do not respect the natural bipartite orientation (black $\rightarrow$ white) of the graph; instead the edges forming a zig-zag path are oriented alternately with, and against the natural orientation.  Note also that the definition of a zig-zag path depends on a choice of embedding for the graph into a Riemann surface: the same abstract graph (a collection of vertices connected by edges) may sometimes be embedded in many topologically inequivalent ways into the same Riemann surface.  The same abstract graph may also be embedded into different spaces; these are both examples of {\it graph isomorphisms}.

We may generalize the notion of zig-zag paths to allow oriented paths on the {\it embedding space} of the graph (the Riemann surface) which cross the edges of the graph along a sequence of edges alternating maximally left, then maximally right, as in the previous paragraph.  This can be thought of as a deformation of a path satisfying the previous definition (which was confined to lie along the edges of the graph); we remark that with this generalization the zig-zag paths define BPS mesons according to the prescription of section \ref{s:spectrum} \cite{Benvenuti:2005ja}.  By deforming the two zig-zag cycles passing through each edge, we can choose their intersection point to lie at a point on the edge (e.g.~at the midpoint).  From now on we use this use this generalized notion of zig-zag path.

\begin{figure}[tbh]
\begin{center}
 \epsfig{file=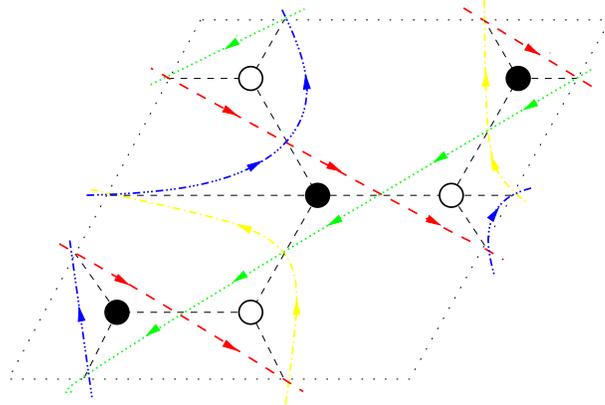}
\end{center}
\caption{The medial graph $\Gamma^m$ of zig-zag paths superpimposed on the brane tiling of del Pezzo 1.\label{f:dp1-zz}}
\end{figure}

The {\it medial graph}, $\Gamma^m$, is the graph formed by the union of all zig-zag paths associated to an underlying graph $\Gamma$, see figure \ref{f:dp1-zz}.  It is 4-valent, with the vertices formed from the intersection of the two zig-zag paths passing through each edge.  Given a bipartite graph, the associated medial graph has the interesting property that each vertex of the bipartite graph is encircled by a face of the medial graph, with either clockwise or anticlockwise orientation correlating with the bipartite colour.  Thus, the medial graph $\Gamma^m$ contains equivalent information about the structure of the underlying bipartite graph $\Gamma$, and we may freely pass from one to the other.

For a general bipartite graph on $T^2$ a zig-zag path may self-intersect, or have zero winding number around the torus.  Neither of these situations arise in the many known brane tilings, so we may raise this to the status of a constraint on the class of bipartite graph that is relevant for the brane tilings\footnote{One way to prove the universality of this constraint would be by establishing that it holds true under Higgsing/partial resolution, since we show in section \ref{s:alga} that it is true for all orbifolds $\C^3/(\Z_n \times \Z_m$), and the brane tiling for any toric singularity can be constructed from such a theory \cite{Hanany:2005ve}.}.

The medial graph is really the fundamental object of the brane tilings; we will see that it plays a key physical role in the mirror geometry.  We note that it has the interesting property that by deforming the zig-zag paths to either meet at the vertices or in the center of the faces of $\Gamma$, one may continuously interpolate between two singular limits: the bipartite graph and its planar dual graph, the planar quiver.

\subsubsection{Knot theory and the Seifert surface}

\begin{figure}[tbh]
\begin{center}
 \epsfig{file=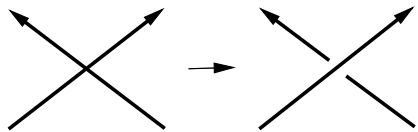}
\end{center}
\caption{Lifting the crossing of two zig-zag paths into knots.\label{f:resolve-knot}}
\end{figure}

The key advantage of introducing the medial graph is that it admits a natural lifting to a thickening of the Riemann surface, e.g.~$T^2 \times I$ where $I = [0,1]$ is a unit interval (we focus for now on the brane tilings on $T^2$; the technique applies to any Riemann surface, and we will make use of this later).  The zig-zag paths of the graph lift to {\it knots} (embeddings of $S^1$) in this 3-manifold, by performing the canonical operation shown in figure \ref{f:resolve-knot}.  The medial graph for a bipartite graph lifts to a {\it completely alternating link}: with respect to the projection onto $T^2$, each oriented knot of the link alternately passes over, then under a link component (technically the medial graph is equivalent to an alternating projection of this alternating link).

This link is topologically very interesting.  A standard tool in singularity theory is the Seifert construction, which produces a Riemann surface whose $S^1$ boundaries are the knots of the link in a 3-manifold.  We now introduce the Seifert construction and study the topology of this Seifert surface.  We will then promote the Seifert surface to a holomorphic curve in a natural way, and it will turn out to be the familiar mirror curve $\Sigma$.

\begin{figure}[tbh]
\begin{center}
 \epsfig{file=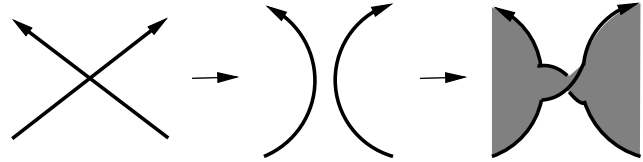}
\end{center}
\caption{Resolving the crossing of two zig-zag paths into Seifert circles, and constructing the Seifert surface.\label{f:resolve-seifert}}
\end{figure}

The first step of the Seifert construction is to apply to the medial graph the operation shown in figure \ref{f:resolve-seifert}.  This resolves the graph into a set of circles, called Seifert circles.  For the brane tilings the Seifert circles enclose the vertices (this follows from the enclosure of the vertices by circles formed from the zig-zag paths, which we have already noted).  It also follows from the nontrivial winding of the zig-zag paths on $T^2$ that these Seifert circles are always non-nested.

\begin{figure}[tbh]
\begin{center}
 \epsfig{file=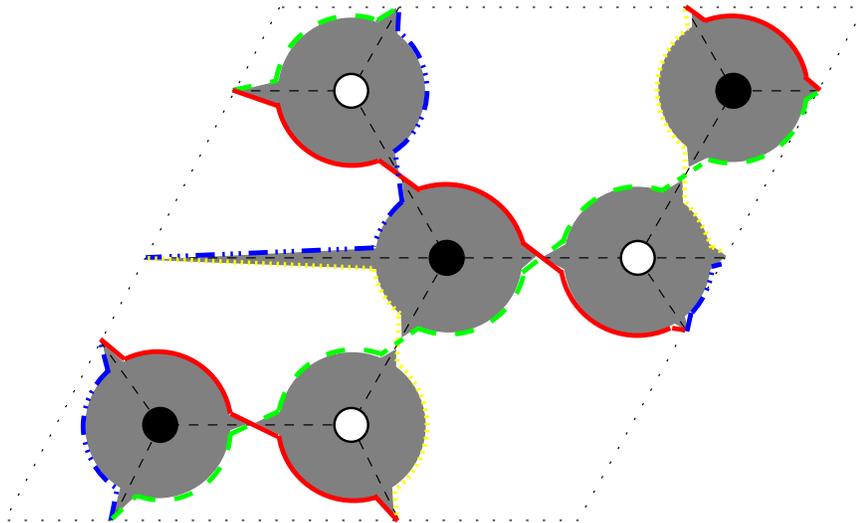}
\end{center}
\caption{The Seifert construction for the del Pezzo 1 brane tiling.  Notice that the boundary circles of the Seifert surface (shaded region) are the zig-zag paths of the bipartite graph embedded in $T^2$ (thick lines), which come from the alternating projection of the link.  The construction induces a graph isomorphic embedding of the bipartite graph into the Seifert surface.\label{f:seifert}}
\end{figure}

To form the Seifert surface, we attach disks to each of the Seifert circles, and then attach a half-twisted strip to each of the zig-zag crossings, as shown in figures \ref{f:resolve-seifert} and \ref{f:seifert}.  By construction, the boundary of this Riemann surface projects to the medial graph on $T^2$, and each boundary component corresponds to a zig-zag path, or a knot in the thickening of this surface.  Topologically we have obtained a Riemann surface whose boundary components are 1-1 with the zig-zag paths on $T^2$.

\begin{figure}[tbh]
\begin{center}
 \epsfig{file=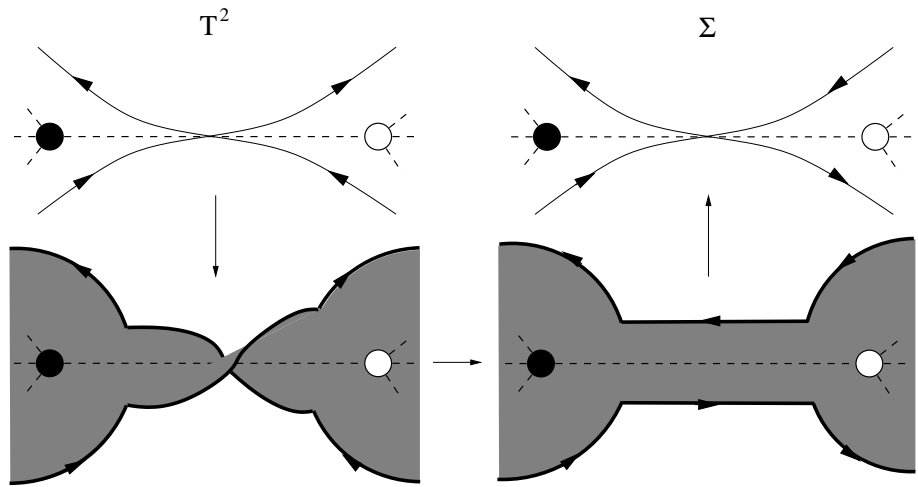}
\end{center}
\caption{Untwisting the twisted strips to produce part of the Seifert surface $\Sigma$ for a bipartite graph on $T^2$.  The zig-zag paths on $T^2$ bound faces of the new (graph isomorphic) embedding of the bipartite graph into $\Sigma$, and similarly the faces on $T^2$ are now the zig-zag paths of the new embedding.\label{f:twist}}
\end{figure}

Note that the Seifert construction induces an isomorphism on the underlying bipartite graph: it is clear from figure \ref{f:seifert} that the bipartite graph of the brane tiling on $T^2$ maps to an embedding of the same abstract graph into the Seifert surface (it has the same vertices and connectivity by edges, but a different notion of ``faces'').  In fact we can be precise about how the embedding changes: the Seifert construction interchanges the role of zig-zag paths and faces, as shown in figure \ref{f:twist}.  Each 4-valent intersection of zig-zag paths of the medial graph on $T^2$ is resolved into a twisted strip; if we untwist each of these strips and contract back to a 4-valent vertex then the former zig-zag paths are now bounding faces, and the former faces are now zig-zag paths with respect to the embedding in the Seifert surface\footnote{At the level of the underlying graph, this operation is called the ``antimap'' \cite{Lins:1980} or the ``Petrie'' map \cite{GraphTheory}.}.

What is the genus of this Riemann surface? It is a standard theorem \cite{Gabai} that the Seifert surface for an alternating projection of an alternating knot has minimal genus (in general it can depend on the projection).  The Euler characteristic is

\begin{equation}
 \chi = s - n
\end{equation}
where $s$ is the number of Seifert circles and $n$ the number of crossings of the projection.  For the brane tilings $s = V$, the number of bipartite vertices, and $n = E$, the number of edges.  Thus, we have for the Seifert surface bounding the link

\begin{equation}
 \chi = 2-2g - h = V - E
\end{equation}
If we glue punctured disks to the $h$ boundaries of the Seifert surface, we obtain a non-compact Riemann surface $\widetilde \Sigma$ without boundary, which has a genus that is yet to be determined.  Note that the bipartite graph of the brane tiling $\Gamma \subset T^2$ is mapped into a graph isomorphic embedding on this new Riemann surface, $\Gamma \subset \widetilde \Sigma$.  In the next section we will show that in fact we may identify $\widetilde \Sigma \simeq \Sigma$, the holomorphic curve of mirror symmetry.  This provides the connection of the brane tilings to the mirror geometry, and we will recover the brane tiling dictionary from the mirror Calabi-Yau and D6-brane system.

\subsection{A closer look at the geometry: am{\oe}b\ae\ and alg\ae}
\label{s:alga}

We may obtain some valuable insights into the geometry of the intersecting D6-brane system by studying two half-dimensional projections of the ambient $(\C^*)^2$ space.  Making use of the polar decomposition $z = r e^{i \theta}$ we have $(\C^*)^2 \simeq T^2 \times \R^2$, and the space admits the two projections by the maps

\begin{eqnarray}
z &\mapsto& \mbox{Log}(|z|)\\\label{e:amoeba}
z &\mapsto& \mbox{Arg}(z)\label{e:alga}
\end{eqnarray}We may also apply these projections to the curve $P(w,z; t)=0 \subset (\C^*)^2$.  The former projection (\ref{e:amoeba}) is called the {\it am{\oe}ba} projection; the latter is called the {\it alga} projection \footnote{We are informed that this was also called the co-am{\oe}ba projection by Passare.}

\begin{figure}[tb]
\begin{center}
 \epsfig{file=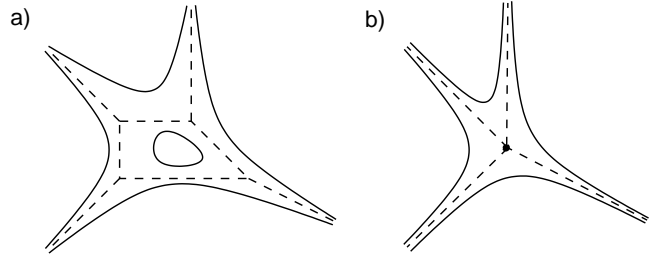}
\end{center}
\caption{The am{\oe}ba of the holomorphic curve $P(w,z) = 1 + \frac{1}{w} + \frac{1}{z} + e^{-s} z - e^{-t} w z$ for the mirror to del Pezzo 1, with dashed lines showing the spine/$(p,q)$ web, a) at a generic Harnack point ($s=3, t = 5$); b) at an isoradial point where $P(w,z) \equiv \det K(w,z)$ with the edge weights determined by the trial R-charges.  The compact oval has degenerated and the complex curve is at a conifold point in the moduli space.\label{f:amoeba}}
\end{figure}
The term ``am{\oe}ba'' comes from the image of Harnack curves under this map, as shown in figure \ref{f:amoeba}.  The am{\oe}ba has a number of ``tentacles'', which are non-compact regions (punctures on the Riemann surface), and it may have zero or more holes depending on the genus of the complex curve and the values of the real moduli.  As the moduli of the Harnack curve are varied, the size of the holes and position of the tentacles may change; these are coordinates on the moduli space of Harnack curves \cite{Kenyon:2003ui}.  At special loci in moduli space the holes shrink to a point.

The locus where all holes of $P(w,z)=0$ have shrunk is the conifold point in the Calabi-Yau moduli space.  Here the genus zero Harnack curve is represented by the spectral curve of an isoradially embedded brane tiling.  If one varies the moduli beyond this limit point, one or more of the holes disappears completely from the projection (the hole has opened up again in the transverse directions to the am{\oe}ba projection).  The curve $P(w,z)=0$ remains a real curve but is no longer Harnack (since it does not have a maximal number of ovals).

It is interesting to note that the Harnack curves have a close relation to the $(p,q)$ webs discussed in section \ref{s:toric}, via the branch of mathematics called {\it tropical geometry} (see \cite{mikhalkin-2000-2,mikhalkin1} for reviews).  Briefly, in the large complex structure limit $t \rightarrow \infty$ together with a rescaling of $(w,z)$ the am{\oe}ba becomes piecewise linear and degenerates onto the $(p,q)$ web.  For finite $t$ the $(p,q)$ web may be embedded in the interior of the am{\oe}ba projection: it forms the so-called ``spine'' of the am{\oe}ba\footnote{Apparently mathematicians do not make good biologists.}.

The alga projection was introduced in this context in \cite{Feng:2005gw}\footnote{The name is intended to evoke a periodic tiling of unicellular organisms on the surface of a pond, as in the infinite cover of $T^2$ by fundamental domains.  Apparently physicists do not make good biologists either.}.  The properties of this projection have not been well studied by mathematicians\footnote{A recent work \cite{vanStraten} proposes to refer to the study of alg{\ae} by the name of ``arctic
 geometry'', after the {\it arg} projection.}, although it is possible to demonstrate some of its properties.

Firstly, we note that the alga projection always admits certain straight-line winding cycles.  These are given by the contours at the punctures of the curve $P(w,z) = 0$.  This follows by considering a certain scaling limit of the curve.  Without loss of generality, by performing an $SL(2,\Z)$ transformation and rescaling we may bring the curve to the form 

\begin{equation}
 P(w,z) = c_1 + c_2 z^{-q} w^p + \sum_i c_i z^{-q_i} w^{p_i} = 0
\end{equation}
where the first two terms form an edge on the boundary of the Newton polygon (we assume for the moment that this side has length 1).  Rescaling

\begin{eqnarray}
z &\mapsto& \lambda^p z\\
w &\mapsto& \lambda^q w
\label{e:scalecstar}
\end{eqnarray}
the curve becomes

\begin{equation}
 P(w,z) = c_1 + c_2 z^{-q} w^p + \sum_i c_i \lambda^{(p,q)\cdot(-q_i,p_i)} z^{-q_i} w^{p_i} = 0
\end{equation}
Since the Newton polygon is convex, in the limit $\lambda \rightarrow \infty$ only the first two terms survive, and the curve becomes

\begin{equation}
 P(w,z) = c_1 + c_2 z^{-q} w^p = 0
\end{equation}
i.e.~in the neighbourhood of a puncture the curve approximates the flat cylinder

\begin{equation}
 z^{-q} w^p = - \frac{c_1}{c_2}
\end{equation}
In particular, it admits the $\C^*$ action (\ref{e:scalecstar}).  Under both the am{\oe}ba and alga projections this cylinder projects to a straight line, with winding number $(p,q)$ on the $T^2$ domain of the alga map.  The offset of these straight lines in the two projections are given by the modulus and argument of $-\frac{c_1}{c_2}$, respectively.  In the general case, when the integer length of a facet of the Newton polygon is $n$, we obtain in the scaling limit a product of $n$ parallel cylinders and similar arguments apply.

The most interesting situation for our purposes is when these straight lines form the boundaries of the alga projection of the curve.  When the alga consists of polygonal regions touching pairwise at their corners and bounded with consistent orientation by straight line winding cycles, we say that the alga projection is ``clean''.
In the clean situation one may also cover the curve $\Sigma$ by patches that are $1:1$ with the polygonal regions of its alga projection.

As we evolve a contour away from the puncture, it must continue to retain the same winding number on $T^2$, but the curve no longer admits the $\C^*$ action and its image under the alga map must deform.  In the clean situation the contour must deform into the interior of the polygonal regions, and there is a limit in which these contours meet pairwise in the interior of each $n$-valent polygon and form an $n$-valent vertex.

Since the curve is $1:1$ with the alga, these project from a decomposition of the curve into punctured disks that meet and glue together along a graph.  But this is the situation we have already encountered in the Seifert construction; indeed, the bipartite graph of the dimer model provides this decomposition of $\Sigma$ into punctured disks.  The brane tiling graph may always be inscribed inside the clean alga projection, and is uniquely defined by it.  It may be considered as the spine of the alga, although it is not known whether it can be recovered in a suitable scaling limit of the curve, as happens for the am{\oe}ba (in the large complex structure limit the alga projection retracts onto the straight line boundary segments\footnote{One might say that our alg{\ae} have exoskeletons, as well as spines.}).  

We may now return to the topological construction of the previous section and compute the genus of the Riemann surface $\widetilde \Sigma$ constructed from the Seifert surface.  The missing data was the number of link components (zig-zag paths on $T^2$), which form the boundary of the Seifert surface.  We have now seen that these correspond to contours enircling the punctures of $\Sigma$, and are counted by the integer perimeter length of the Newton polygon.  According to Pick's theorem \cite{Pick},

\begin{equation}
 2 \mbox{Area}(\Delta) = 2 I + \mbox{Perimeter}(\Delta) - 2
\label{e:pick}
\end{equation}
where $A = \mbox{Area}(\Delta)$, $P = \mbox{Perimeter}(\Delta)$, and $I$ is the number of interior lattice points of $\Delta$.  By the correspondence to the $(p,q)$-webs (which is a theorem of tropical geometry), $I$ is equal to the genus of the curve $P(w,z)=0$.  The left-hand side of (\ref{e:pick}) is the number of triangles in a complete unimodular triangulation of $\Delta$, and is also equal to the number of gauge groups of the quiver theory.  Therefore, we have for the genus $g$ of the Seifert surface,

\begin{eqnarray}
2-2g &=& V - E + P\\
-2g &=& V - E + F - 2 I\\
g &=& I
\end{eqnarray}
where we have used $V-E+F=0$, the Euler characteristic of the brane tiling on $T^2$.  Thus, after we glue a punctured disk to each of the boundary knots of the Seifert surface, that Riemann surface is homeomorphic to the mirror holomorphic curve $\Sigma$.

Thus, we have an algebraic explanation for the topological construction of the previous section.  The zig-zag paths on $T^2$, which have $(p,q)$ winding, correspond to contours encircling the punctures on $\Sigma$.  The topological Riemann surface $\widetilde \Sigma$ of the Seifert construction may be naturally identified with the holomorphic curve $\Sigma$.


\begin{figure}[tb]
\begin{center}
 \epsfig{file=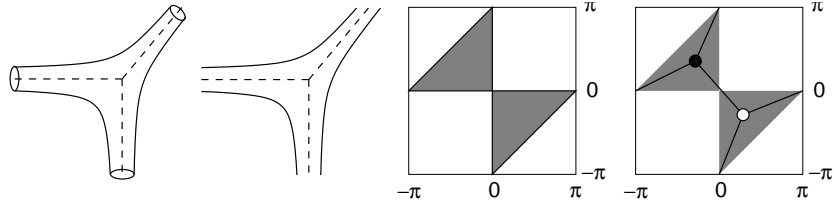}
\end{center}
\caption{a) The holomorphic curve $P(w,z; t) = 1 + w + z \subset (\C^*)^2$ for the mirror to $\C^3$; its amoeba projection to $\R^2$ (with spine shown by dashed lines); and its alga projection to $T^2$; the unique bipartite graph induced by the alga projection in this case.\label{f:c3}}
\end{figure}
We now show that for Abelian orbifolds of $\C^3$ the orbifold point in moduli space produces a clean alga.  This follows from the alga projection of the curve for $\C^3$, shown in figure \ref{f:c3}.  The curve $P(w,z; t) = 1 + w + z$ (corresponding to the mirror to $\C^3$) is a $2:1$ projection over its amoeba, with the boundary real and the fibre over the interior two points related by complex conjugation.  The top and bottom sheets of this projection map to the two triangles of the alga: the straight line segment in the $(p,q)$ direction projects from a contour near one of the 3 punctures which winds halfway around it.  The corner point of the triangle is real and projects from the path along one of the real boundary components of the amoeba from one puncture to the next.

\begin{figure}[tbh]
\begin{center}
 \epsfig{file=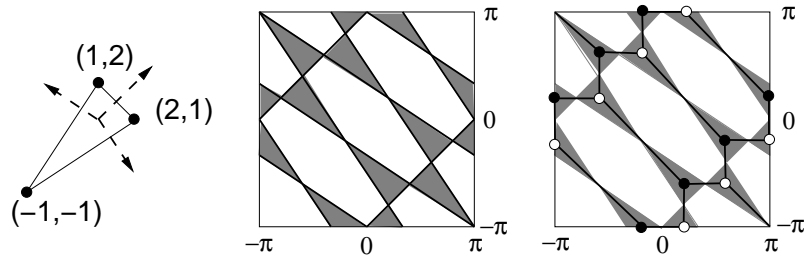}
\end{center}
\caption{The Newton polygon of $P(w,z; t) = \frac{1}{zw} + z^2 w + w z^2 \subset (\C^*)^2$ for the mirror to $\C^3/\Z_5$ with action $(x,y,z) \sim (\lambda x, \lambda^2 y, \lambda^2 z)$, $\lambda^5 = 1$; its alga projection to $T^2$; the unique bipartite graph induced by the alga projection.\label{f:c3z5}}
\end{figure}

The result for arbitrary orbifolds $\C^3/\Gamma$,where $\Gamma \simeq \Z_n$ or $\Z_n \times \Z_m$, follows by applying an appropriate $GL(2,\Z)$ transformation on the toric diagram, i.e.~mapping

\begin{eqnarray}
 z &\mapsto& z^a w^b\nonumber\\
w &\mapsto& z^c w^d
\label{e:gl2}
\end{eqnarray}
It is easy to see that the resulting curve (with all other lower-order monomials set to 0) is at the orbifold point in the complex structure moduli space.  For example, the curve

\begin{equation}
 P(w,z; t) = 1 + z^n + w^m
\end{equation}
corresponds to the orbifold $\C^3/(\Z_n \times \Z_m)$, and indeed admits the orbifold symmetry generated by phase rotations of $(z, w)$ by $(n,m)$ roots of unity respectively.  The action on the alga projection is induced by the map (\ref{e:gl2}) and amounts to tiling $n \times m$ copies of the alga of $\C^3$ inside the $T^2$ of the alga projection.  Diagonal orbifolds $\C^3/\Gamma$ may also be treated similarly, as in figure \ref{f:c3z5}.
A similar construction also holds true for orbifolds of the conifold at the point $P(w,z; t) = 1 + z + w - zw$; this produces a clean alga which is tiled in a checkerboard pattern.

At a generic point in complex structure moduli space, the alga projection of a curve $P(w,z; t)=0$ is not clean.  e.g.~the projection does not consist of polygonal regions touching at corners, but also contains fold singularities.  The only known technique for plotting the alga in this situation is to use Monte Carlo sampling of points on the curve.  However, even in this situation, the alga projection must always admit the straight-line winding paths coming from contours at the punctures; they just typically no longer form the boundary of the alga and may lie in the interior.

When the alga is clean then it uniquely defines a brane tiling: the straight line boundary winding cycles of the alga are the zig-zag paths, whose winding number agrees by construction with the primitive normals to the toric diagram; and they enclose regions whose boundary has index $\pm 1$ on $T^2$ and has alternating sign between adjacent regions.  However, it is important to keep in mind that we have not identified the particular representative winding cycles that correspond to the special Lagrangian D6-branes, and in general they will be deformations of the straight-line cycles into the interior of the curve and its alga projection.

Recall that the orbifold point is where the fractional branes of the quiver theory are mutually BPS, so this is where we expect the brane tiling to correctly describe the physics of these branes.  The central charge of the special Lagrangian cycles is given by the period integral of the holomorphic 3-form

\begin{equation}
 \Omega = \frac{dw\ dz\ du}{w z u}
\end{equation}
which reduces to the period integral
\begin{equation}
 \oint \log(w) \frac{dz}{z}
\end{equation}
on contours of $P(w,z;t)=0$ \cite{Imamura:2007dc}.  Thus, the central charge of the fractional D6-branes are given by linear combinations of the boundary coefficients of $P(w,z;t)$, and they are expected to have equal phases at the orbifold point.  Indeed, it is encouraging (but not {\it a priori} required) that we indeed recover the correct brane tiling by a projection of the mirror geometry at the orbifold point.  The fact that this holds precisely at the orbifold point for all theories $\C^3/\Gamma$ and $\{\mbox{Conifold}\}/\Gamma$, where $\Gamma \simeq \Z_n$ or $\Z_n \times \Z_m$, suggests that a similar statement may also hold for more general toric Calabi-Yaus; the full generality of this result is currently unknown.

\begin{figure}[tbh]
\begin{center}
 \epsfig{file=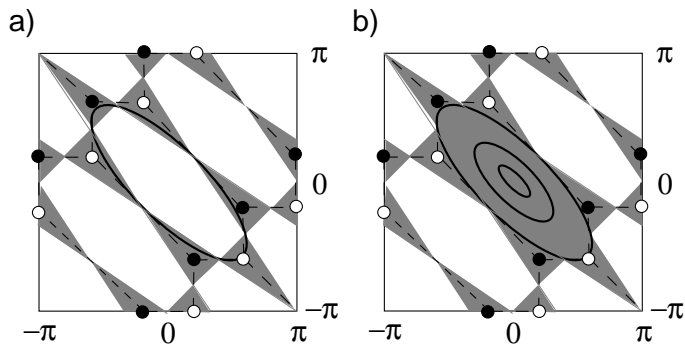}
\end{center}
\caption{a) The vanishing cycle corresponding to the intersection of the D6-brane with the curve $P(w,z; t)=0$.  b) The shaded region shows the face of the brane tiling which fills in along the straight line path from $W=0$ to $W=W_*$ (here we have chosen 
$W_* = \frac{2}{3^{3/5}} + 3^{2/5}$, one of the 5 critical points of $P(w,z; t) = \frac{1}{zw} + z^2 w + w z^2$).  At intermediate points the alga projection contains a hole bounded by the black curves, which vanishes completely in the limit $W \rightarrow W_*$; this is consistent with the images of the vanishing cycle spanning the lift of this disk in $(\C^*)^2$.\label{f:vanishing}}
\end{figure}

We may use the alga projection to provide further evidence for the D6-brane construction of the brane tiling.  The faces of the brane tiling should correspond to the disks that project from the D6-branes.  The boundary of these disks must lie in the curve above $W=0$, and indeed we may inscribe the graph of the brane tiling within the alga of the curve $P(w,z; t)=0$.  Moreover, the interior of the face should be filled out by the image of this 1-cycle that vanishes at a critical point $P(w,z; t) = W_*$.  Indeed, if we look at the alga projection of the curve along the straight line path from the origin to a critical point, the corresponding face of the alga projection fills in; see figure \ref{f:vanishing}.  This is consistent with the face being the image of the disk bounded by the vanishing cycle.

\subsection{Brane tilings and mirror symmetry}

We now have the following picture of the geometry of the brane tilings.  The mirror Riemann surface $\Sigma = \{P(w,z; t) = 0\} \subset (\C^*)^2$ intersects with the $T^2$ of the brane tiling, which is wound around the angular directions of $(\C^*)^2$.  These two surfaces intersect along the medial graph $\Gamma^m$, which may therefore be thought of as embedded into either surface, $\Sigma$ or $T^2$.

The two surfaces are related by the Seifert construction: with respect to either embedding, deform the zigzag paths of the embedding in a normal direction to produce an alternating link.  This describes the geometry away from the intersection locus $\Sigma \cap T^2$.  For the embedding $\Gamma^m \subset T^2$ this deformation corresponds to pushing the zigzag paths (which each encircle a puncture of the $\Sigma$) in the direction of this puncture.  A zigzag path lifts to a $(p,q)$ torus knot with respect to the angular directions of $(\C^*)^2$, and encircles the $(p,q)$ cylindrical end of the curve $\Sigma$.  Gluing in a punctured disk to each of the boundaries of the Seifert surface for the embedding in $T^2$, we obtain a Riemann surface which is homeomorphic to $\Sigma$.

Conversely, if we consider at the graph $\Gamma^m$ as embedded in $\Sigma$, then the zig-zag paths are the $F$ winding countours corresponding to the intersection of the D6-branes with $\Sigma$.  Deforming normally with respect to $\Sigma$ we obtain a different alternating link, whose Seifert surface is topologically $T^2$ with $F$ holes.  Gluing a disk (unpunctured) to these holes we fill out $T^2$.


\subsubsection{Recovering the dictionary}

We may now confirm the ``dictionary'' of the brane tilings that was proposed in figure \ref{t:dictionary}.  The D6-branes in the mirror Calabi-Yau project to disks in $(\C^*)^2$ whose boundaries are zig-zag 1-cycles on the medial graph as embedded into $\Sigma$.  By the Seifert construction, they indeed correspond to the faces of the medial graph as embedded into $T^2$.  Points where two of these 1-cycles intersect on $\Sigma$ are also points on $T^2$ where two faces touch; at these points in $(\C^*)^2$ a massless open string may localize to the intersection point of the two D6-branes, producing an $\N=1$ chiral multiplet in the bifundamental representation of the gauge groups of the two D6-branes.  For some geometries a D6-brane may self-intersect; then the chiral multiplet is in the adjoint representation.

The remaining piece of the dictionary is to understand the generation of superpotential terms.  In string theory they are generated by open string worldsheet instantons ending on the D6-branes.  In the field theory limit ($g_s \rightarrow 0$) only the holomorphic disk instantons contribute, with weight $e^{-A}$ where $A$ is the area of the holomorphic disk.  The only compact holomorphic disks bounded by the D6-branes on $\Sigma$  are the contours encircling the vertices of the medial graph; all other contours enclose a puncture of the curve, so the worldsheets are non-compact and give a vanishing contribution to the superpotential.

From the point of view of the disk worldsheet, the D-brane intersection points correspond to boundary condition changing operators inserted on the boundary of the disk.  In spacetime these operators correspond to the $\N=1$ chiral multiplets; therefore the contribution of the disk instanton is

\begin{equation}
 W = \sum_{V \in \mbox{vertices}} e^{-A(V)} \Tr \prod_{e_i \in V} X_i
\end{equation}
where $A(V)$ is the area of the disk encircling the bipartite vertex $V$, which acts as a superpotential coupling; $e_i$ are the edges adjacent to this vertex, in the appropriate (clockwise or anticlockwise) ordering compatible with the orientation of the disk; and $X_i$ the chiral multiplet of the quiver theory.  This is precisely the required contribution to produce the superpotential of the quiver theory.

We may now also exhibit the $T^3$ of mirror symmetry which is spanned (in homology) by the D6-branes.  We have shown that the faces of the $T^2$ correspond to the projection of the D6-branes.  The projected direction is the $S^1$ in the $W=-uv$ fibre of the Calabi-Yau.  Recall that this $S^1$ is finite away from $W=0$ and vanishes at $W=0$; but the latter is the locus of the medial graph, which is where the D6-branes intersect $\Sigma$.  Therefore this $S^1$ fibres over the $T^2$ of the brane tiling so that it vanishes along the medial graph and is finite elsewhere (in particular, it is finite over the interior of the disks, so we recover the $S^3$ topology of the D6-branes).  The other point of interest is that part of the world-volume of the $T^3$ (the neighbourhood of the vertices) is spanned by the disk instantons.

To summarize the results of this section, for any brane tiling the procedure we have described produces a collection of winding cycles on the mirror Riemann surface $P(w,z; t) = 0$ whose intersections precisely reproduce the data of the quiver theory, and which also reproduce the superpotential terms, via the holomorphic disk instantons bounded by these contours.   Moreover, by using the alga projection it can be explicitly checked that the faces of the brane tiling correspond to the alga projection of the D6-branes.  We take this as strong evidence that in the general case these are the correct (vanishing) homology 1-cycles that project from the Lagrangian 3-spheres; this equality may also be confirmed directly in the known examples where this homology has been computed using other methods \cite{Hanany:2001py,Orlov}. 

\subsubsection{The T-dual NS5-D5 system}

As we have discussed in detail, there are two sets of paths on $\Gamma \subset \Sigma$ that define the brane tilings, the zig-zags and faces.  We have given a physical interpretation to the zig-zag paths on $\Sigma$: they are the intersection of the D6-branes with the curve $\Sigma$, and bound disks which fill out the faces of the brane tiling on $T^2$.  The curve $\Sigma$ itself may be constructed by attaching holomorphic disks at the vertices of $\Gamma$, and punctured holomorphic disks at the faces.  Physically, the former are spanned by open string worldsheets and give rise to superpotential terms in the gauge theory, but what about the latter?  In the mirror picture the curve $\Sigma$ itself is not wrapped by anything, so the punctured disks seem to have no physical meaning.

However, there exists a dual picture in which they do take on physical meaning.  T-dualizing on the circle in the $W = -uv$ fibre, one obtains a NS5-brane wrapping the locus where the fibre degenerates, i.e.~wrapping the entire curve $P(w,z; t)=0$.  The D6-branes wrap the fibre, and dualize into D5-branes.  Thus, we obtain an intersecting NS5-D5 system that is also characterized by the discussions of the previous section, and this is another equivalent way to obtain the brane tilings in string theory.

This picture is very similar to the old ``brane box'' models for orbifolds of the conifold \cite{Hanany:1997tb,Aganagic:1999fe}.  There were earlier attempts to generalize the brane box models to orbifolds of $\C^3$ \cite{Hanany:1997tb,Hanany:1998it}, but these involved {\it ad hoc} rules that did not have a clear physical justification.  In fact, the brane tilings provide the resolution to this puzzle: for orbifolds of $\C^3$ the brane tiling is a tiling by regular hexagons, which recovers the rules of \cite{Hanany:1997tb,Hanany:1998it}.

This NS5-D5 version of the brane tilings has been systematically exploited by \cite{Imamura:2006db,Imamura:2006ie,Imamura:2007dc} to study various gauge-theoretical properties of the brane tilings such as anomalies and marginal deformations.

\subsection{A proposal for the construction of the toric quiver theories}

We are now ready to propose a general method for constructing the quiver gauge theory on the world-volume of a stack of D3-branes located at a toric singularity.

At the level of topology there is a natural proposal consistent with all of the properties we have seen so far: take a set of $(p,q)$ winding cycles corresponding to the primitive outward-pointing normal vectors to the boundary of the toric diagram (counted with multiplicity given by the integer length of the sides of the polygon).  Find a non-self-intersecting arrangement of these winding cycles that form the medial graph of a bipartite graph.  i.e.~proceeding along each winding cycle we encounter other cycles that alternately approach from the left and the right, and the winding cycles bound faces that touch at the 4-valent vertices and alternate with opposite boundary orientation.  This is a refinement of the proposal of \cite{Hanany:2005ss}\footnote{It is often the case that there may be more than one brane tiling for a given Calabi-Yau that are consistent with these rules; these are in fact related by a version of Seiberg duality \cite{Franco:2005rj} and are expected to flow to the same IR fixed point.  However, there is a constraint from the principle of $a$-maximization: in examples one finds that there is only one brane tiling that admits a physically sensible $a$-maximal charge assignment, and the other Seiberg-dual theories have fields with R-charges that are driven to $0$, which is interpreted as unphysical.  In terms of the isoradial embedding, they correspond to embeddings that degenerate onto the brane tiling of another phase.  It appears that typically only a single theory is compatible with $a$-maximization and the other Seiberg dual theories must undergo RG flow to this one.}.

Equivalently, in terms of knots, we are instructed to find an arrangement of $(p,q)$ torus knots that form a completely alternating link, and take an alternating projection to the torus.  This projection recovers the medial graph and thence the bipartite graph.

However, we can say more by using the geometrical results of section \ref{s:alga}.  The winding cycles on $T^2$ cannot be arbitrary, but must lie on the curve $P(w,z; t)=0$.  Therefore they must be contained in the alga projection of this curve.  As we have seen, it is often the case that by simply taking the alga projection of the curve $P(w,z; t)=0$ at the orbifold point in moduli space, the required $(p,q)$ winding cycles are homotopic to the boundary of the alga, and the bipartite graph of a brane tiling for this space may be read off immediately and without ambiguity.  The full generality of this result is currently unknown.

\section{Further work}
\label{s:further}

The dimer models were first introduced into string theory in \cite{Okounkov:2003sp} as a proposed statistical model of closed topological strings on toric Calabi-Yau manifolds.  While it is manifest that the dimer models for $\C^3$ indeed computes the topological string partition function to all orders in $g_s$ (by an explicit isomorphism to the microstates of the melting crystal model), for other Calabi-Yau manifolds the correspondence (at least as currently understood) is only true at genus 0 in the string expansion.  This may be easily checked by comparing the expansion of the crystal partition function of \cite{Okounkov:2003sp} for e.g.~the conifold, with the dimer partition function; one finds that they do not agree.

Nevertheless, the fact that there is even agreement between these two partition functions at string tree level (see e.g.~\cite{Stienstra:2005a}) is intriguing (and non-obvious, given the non-isomorphism of microstates).  Moreover, the dimer models used in \cite{Okounkov:2003sp} for a given toric Calabi-Yau are the same graphs as in the present work on brane tilings (the difference is that the former works on the infinite cover of $T^2$ and imposes certain boundary conditions).  As we have seen, the brane tilings are associated to the {\it open} string theory in the $g_s \rightarrow 0$ limit.  One attempt to reconcile these two points of view was \cite{Heckman:2006sk}.  Another line of approach is suggested by the role of the infinite cover of $T^2$ in enumerating the BPS spectrum of the gauge theory.  We suspect there is a deeper connection between these various points of view that remains to be uncovered.

The recent work of \cite{Dijkgraaf:2007yr} promotes yet more connections between string theory and dimer models, making use of more results from the long history of dimer models in mathematical physics, and related subjects.  They argue for the equivalence between the dimer model of a brane tiling, and a free fermion propagating on the same graph.   Of particular interest is that a key role is played by the loops formed by differences of perfect matchings, which provide a loop expansion of the fermion determinant.  As we have elucidated in this review, these loops also define the $U(1)$ symmetries of the gauge theory.  A formal analogy with knot homologies is noted, and indeed we have seen that these loops are associated to knots in the mirror geometry.

We also remark that matrix factorizations have played a role in the study of knot homologies \cite{Gukov:2005qp}, and there are natural matrix factorizations defined by the dimer models, via the identity

\begin{equation}
 \det K \equiv K . adj(K) = adj(K) . K
\end{equation}
where $K$ is the Kasteleyn matrix, and $adj(K)$ is the classical adjugate matrix, the transpose of the matrix of cofactors.  It remains to be seen whether this observation has any interesting roles to play\footnote{One may use this family of matrix factorizations to define B-type D-branes in the Hori-Vafa mirror theory.  However, their spectrum is only nontrivial at singular points in both the closed and open string moduli space.  In the geometrical mirror picture, the spectral curve $\det K(w,z) = 0$ must be completely degenerate, and the open string moduli tuned so that the D-branes are located at the singular point(s).  However since the bulk world-sheet theory is singular, the physical interpretation of this result is unclear \cite{KennawayHori}.}.

Other mathematical work on brane tilings \cite{Ueda:2006jn,Ueda:2006wy,Ueda:2007xz} has explored their role in the homological mirror symmetry conjecture.  We hope that some of these ideas will prove useful to mathematicians in this very active area of mathematical research.

Various authors have explored the extension of the brane tiling techniques to study various conformal, non-conformal and non-supersymmetric deformations of the quiver gauge theories \cite{Franco:2005zu,Butti:2005sw,Brini:2006ej,Garcia-Etxebarria:2006aq,Butti:2006hc,Butti:2006nk,Garcia-Etxebarria:2007vh}, as well as possible applications to phenomological model-building \cite{Garcia-Etxebarria:2006rw}.

Another interesting recent development has been the extension of the two-dimensional dimer models to a three dimensional torus, to study three-dimensional CFTs via compactifications of M-theory on a toric Sasaki-Einstein $7$-manifold and the $AdS_4/CFT_3$ correspondence \cite{Lee:2006hw,Lee:2007kv,Kim:2007ic}.  The dual CFTs remain poorly understood, but there has been some progress in understanding topological aspects by making use of the 3-dimensional analogue of the brane tilings.

The analogy noted in section \ref{s:spectrum} that interprets the scalar BPS spectrum as open and closed strings propagating on the brane tiling is intriguing, and it seems to also provide a useful framework for the study of the spin chain for the 1-loop dilatation operator of the gauge theory \cite{KennawayWIP} (see also \cite{Benvenuti:2005cz,Sadri:2005gi}).  It would be interesting to study this ``string theory'' in more detail.

{\bf Acknowledgements:} The author is supported by NSERC.  It is a pleasure to thank my collaborators Bo Feng, Sebastian Franco, Amihay Hanany, Yang-Hui He, Cumrun Vafa, David Vegh and Brian Wecht; and to acknowledge stimulating discussions with Chris Herzog, Kentaro Hori, Amer Iqbal, Richard Kenyon, Grisha Mikhalkin, Andrei Okounkov, Joan Simon, Jan Stienstra, Alessandro Tomasiello and Johannes Walcher.

\newpage
\bibliographystyle{JHEP}
\bibliography{kk.bib}

\begin{figure}[p]
\begin{center}
 \epsfig{file=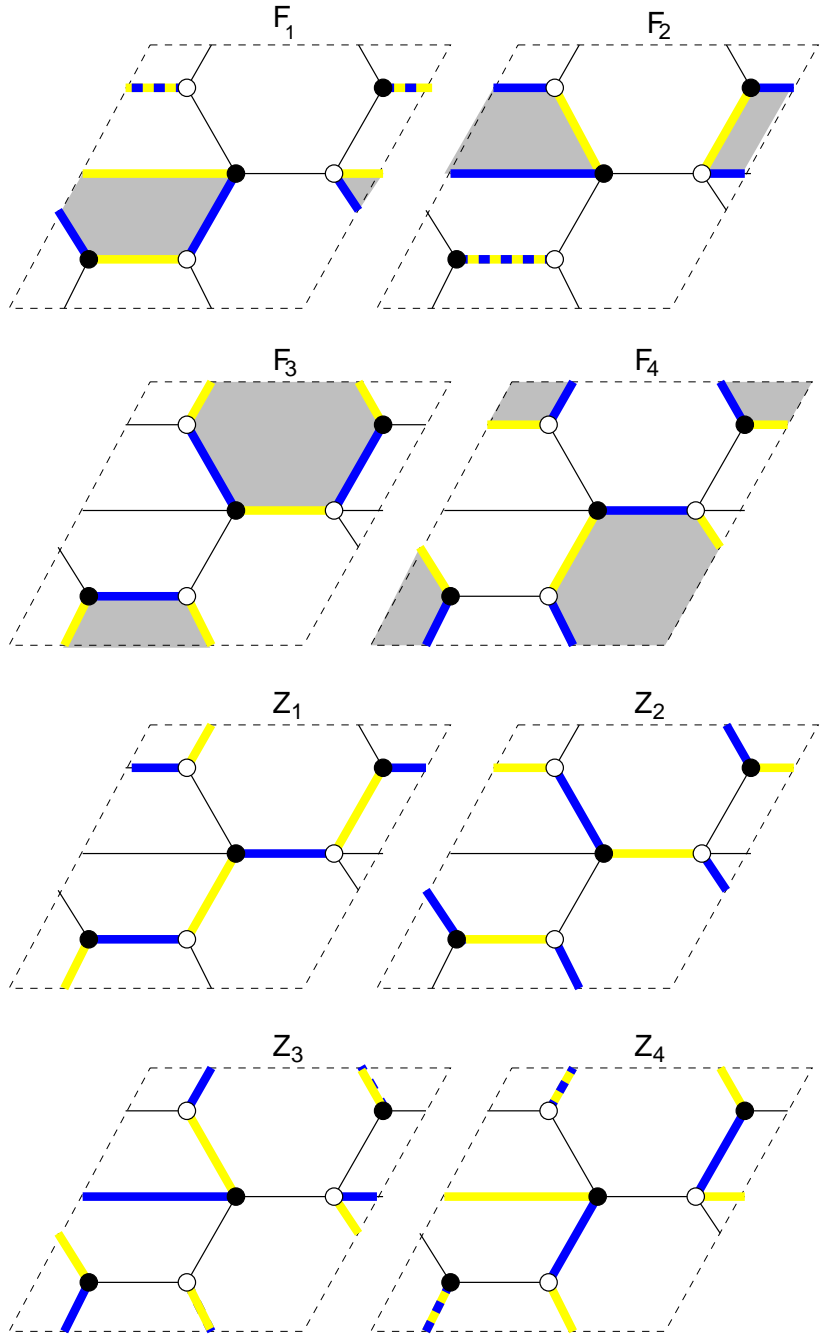,height=7in}
\end{center}
 \caption{The 4 face symmetries and 4 zig-zag symmetries of the del Pezzo 1 quiver formed by differences $m_i - m_j$ of matchings, where $m_i$ are the light-coloured edges, and $m_j$ the dark-coloured edges.  The dashed lines correspond to edges present in both matchings, which cancel out in the difference.\label{f:dp1-sym}}
\end{figure}

\end{document}